\documentclass[aps,prl,twocolumn,amsmath,amssymb,nofootinbib,
longbibliography]{revtex4-2}

\usepackage{graphicx}
\usepackage{dcolumn}
\usepackage{bm}
\usepackage{xcolor}
\usepackage{comment}

\AtBeginEnvironment{pmatrix}{\setlength{\arraycolsep}{-6pt}}

\usepackage[colorlinks=true,bookmarks=false,allcolors=blue]{hyperref} 

\newcommand{\fla}[1]{\begin{flalign}#1\end{flalign}}
\newcommand{\Ain}{A_{\text{in}}}
\newcommand{\Aout}{A_{\text{out}}}
\newcommand{\Bin}{B_{\text{in}}}
\newcommand{\Sin}{\boldsymbol{\Sigma}_{\text{in}}}
\newcommand{\nbar}{\bar{n}}

\newcommand{\Ipair}{\mathcal{I}_{\text{pair}}}

\def\mean#1{\langle#1\rangle}
\newcommand{\Sout}{\boldsymbol{\Sigma}_{\text{out}}}

\begin{document}


\title{Critical quantum metrology in the output of an optical parametric oscillator}

\author{Polina Blinova}
\author{Kai Wang}
\email{k.wang@mcgill.ca}
\affiliation{Department of Physics and McGill Quantum Centre, McGill University, 3600 rue University, Montréal, Québec H3A 2T8, Canada.}

\date{September 14, 2026}

\begin{abstract}
Many quantum sensors infer parameters from continuously emitted fields, yet in the presence of unmonitored loss, it is difficult to determine how much critical enhancement survives in the accessible output. In this work, we calculate the full frequency-resolved output quantum Fisher information (QFI) of a stationary vacuum-seeded optical parametric oscillator for cavity-detuning estimation. We show that, for any fixed nonzero unmonitored loss, the QFI of correlated sideband pairs grows quadratically with the mean intracavity photon number near threshold, but only within a spectral window that narrows inversely with that number; consequently, the spectrum-integrated output QFI rate scales asymptotically only linearly. We identify a far-detuned, near-threshold regime in which the monitored output alone asymptotically approaches the loss-imposed upper bound on the joint cavity–output QFI. 
We further determine the operating points that maximize the output QFI rate per intracavity photon and introduce a frequency-resolved homodyne strategy that maximizes the Fisher information obtainable from a single record. Our results establish an operational connection between critical enhancement and metrological information accessible in the output field, while providing concrete guidance for the design and readout of dissipative parametric sensors.
\end{abstract}

\maketitle

A broad class of sensors operates by driving an open dissipative system and continuously monitoring its emitted field, from gravitational-wave detectors~\cite{aasi2013enhanced,caves1981quantum,ligo2011gravitational} to optomechanical sensors~\cite{aspelmeyer2014cavity}. Information about the sensed parameter is then encoded in both the internal state and the outgoing field, with the attainable sensitivity determined by the quantum Fisher information (QFI) of the accessible state~\cite{degen2017quantum,giovannetti2011advances}. While the QFI of the internal state bounds measurements on the sensor alone, additional information escapes into the emitted field, motivating studies of the QFI carried by the full environment and strategies to retrieve it~\cite{midha2025metrology,yokomizo2026asymptotic,yang2023efficient}. In practice, optical loss leaves only part of the emitted field experimentally accessible. For output-based sensing, the relevant figure of merit is then the QFI of the monitored output, rather than that of the full system-environment state~\cite{macieszczak2016dynamical,gammelmark2014fisher}. Computing the monitored-output QFI after tracing out unmonitored channels is generally nontrivial, motivating computational approaches~\cite{albarelli2026efficient,yang2026quantum,khan2025tensor} or analysis of spectrally filtered output modes~\cite{vivas2026quantum}.

Parametric oscillators offer two distinct features relevant to emitted-field sensing: squeezing can reduce quadrature noise~\cite{caves1981quantum}, whereas exceptional points (EPs) can enhance certain responses to parameter perturbations~\cite{wiersig2014enhancing,wiersig2016sensors,chen2017exceptional,hodaei2017enhanced}. Neither feature alone determines estimation precision, whereas QFI accounts for displacement, squeezing, thermal noise, and correlations within a common framework~\cite{gaiba2009squeezed,pinel2012ultimate,pinel2013quantum,jiang2014quantum,vsafranek2015quantum,vsafranek2019estimation}. EP-enhanced coherent scattering~\cite{wiersig2026fundamental}, EP sensing combined with squeezing~\cite{roy2021nondissipative,wang2026squeezing}, and QFI scaling at EPs~\cite{liu2024scaling} have all been explored. Yet EPs do not automatically improve precision once mode coalescence and noise are included~\cite{langbein2018no,wang2020petermann,chen2019sensitivity}, and reciprocal sensors obey the same fundamental bounds at or away from the EP~\cite{lau2018fundamental}. A distinct route is critical metrology, which seeks to exploit the divergent susceptibility near a phase transition~\cite{garbe2020critical,chu2021dynamic,mihailescu2026critical,zanardi2008quantum}. In open quantum-optical systems, dissipative criticality occurs when the slowest relaxation rate vanishes~\cite{minganti2018spectral}, such as at the parametric-oscillation threshold, where QFI can be strongly enhanced~\cite{zhang2019quantum,anderson2023clarification}. Dissipation and critical slowing down, however, constrain the attainable metrological gain~\cite{gietka2021adiabatic,rams2018limits}, raising the question of whether the enhancement persists as a sustained information rate. Prior critical sensing studies with parametric oscillators focused on the intracavity state~\cite{alushi2024optimality,di2023critical,chen2024critical}, specific output observables~\cite{beaulieu2025criticality}, or the joint system-output QFI~\cite{ilias2022criticality,gualrecki2025time,gorecki2025interplay}. Very recent replica-based calculations have also numerically addressed the output QFI of an optical parametric oscillator under inefficient monitoring~\cite{albarelli2026efficient}, while leaving open an analytical characterization of its critical information rate and the conditions for saturation of the loss-imposed bound. Two questions therefore remain: how much of the critical enhancement survives in the monitored radiation when some loss channels are inaccessible, and under what operating conditions does the monitored output contain essentially all of the QFI available in the joint cavity–output state?

In this Letter, we answer these questions using the optical parametric oscillator (OPO), a minimal model for a broad class of parametrically driven bosonic sensors, including optical~\cite{ou2020quantum} and superconducting parametric amplifiers~\cite{wustmann2013parametric}, Kerr resonators~\cite{guo2024quantum}, and parametrically driven mechanical systems~\cite{levitan2016optomechanics}. We first derive the frequency-resolved QFI of its full stationary output and show that any fixed nonzero unmonitored loss confines the quadratic enhancement of correlated sideband pairs to a narrowing spectral window, leaving the spectrum-integrated QFI rate linear in intracavity occupation; in the fully monitored limit, the integrated rate instead scales cubically. We then identify two distinct near-threshold limits—large detuning and strong overcoupling—in which the monitored output asymptotically saturates the loss-imposed upper bound on the joint cavity–output QFI.
Consequently, the fraction of the joint QFI carried by the monitored output tends to unity.
Finally, at finite photon number, we show that the output-optimal operating points generally differ from the intracavity-QFI optimum, the EP, and the points of maximal squeezing or antisqueezing, and propose a homodyne measurement scheme that maximizes the classical Fisher information over frequency-dependent homodyne quadrature angles from a single record.

\begin{figure}[!t]
    \centering
    \includegraphics[width=1\linewidth]{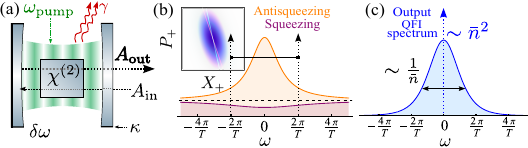}
    \vspace{-0.6cm}
    \caption{OPO output sensing. (a) Illustrative $\chi^{(2)}$ realization of a single-mode OPO pumped at $\omega_{\text{pump}}$ and detuned by $\delta\omega$, with intrinsic loss $\gamma$, output coupling $\kappa$, vacuum input $\Ain$, and output $\Aout$. (b) The output shows frequency-dependent squeezing and antisqueezing; each sideband pair forms a two-mode Gaussian state that changes under a perturbation of $\delta\omega$ (inset). (c) The QFI spectrum grows as $\nbar^2$ near threshold while narrowing as $1/\nbar$, and its integral gives output QFI rate for estimating $\delta\omega$.}
    \label{fig1}
\end{figure}

\textit{OPO model and output QFI.} We study the single-mode degenerate OPO shown in Fig.~\hyperref[fig1]{1(a)}, whose input-output dynamics are given by~\cite{Gardiner1985}
\fla{\begin{pmatrix}
\dot{a} \\ \dot{a}^\dag 
\end{pmatrix}
=\mathcal{E} \begin{pmatrix}
    a \\ a^\dag 
\end{pmatrix}
+\sqrt{\kappa} \begin{pmatrix}
    \Ain \\ \Ain^\dag 
\end{pmatrix} + \sqrt{\gamma} \begin{pmatrix}
    \Bin \\ \Bin^\dag
\end{pmatrix},}
where  $\mathcal{E}=\left(\begin{array}{cc}
     -\frac{\kappa+\gamma}{2}-i\delta\omega& \xi  \\
     \xi & -\frac{\kappa+\gamma}{2}+i\delta\omega
\end{array}\right)$ with $\Gamma=\kappa+\gamma$. Here, $\delta \omega\ge0$ denotes the cavity-mode detuning in the effective rotating-frame model. For the illustrative $\chi^{(2)}$ realization shown in Fig.~\hyperref[fig1]{1(a)}, the frame rotates at $\omega_{\text{pump}}/2$ with $\omega_{\text{pump}}$ being the optical pump's angular frequency. The parametric drive strength $\xi>0$ is assumed real. The cavity couples at rate $\kappa$ to the monitored input-output channel $\Ain$ and at rate $\gamma$ to the unmonitored intrinsic loss bath $\Bin$, which accounts for undetected photons. The non-Hermitian dynamical matrix $\mathcal E$~\cite{wang2019non} has a second-order EP at $\xi=|\delta\omega|$, where its eigenvalues $\lambda_\pm=-\Gamma/2\pm\sqrt{\xi^2-\delta\omega^2}$ coalesce, while the parametric oscillation threshold occurs at $\xi=\sqrt{\delta\omega^2+\Gamma^2/4}\equiv \xi_c$.

We estimate $\delta\omega$ from the stationary monitored output $\Aout=\Ain-\sqrt{\kappa}a$. With vacuum in both external baths, the output is a zero-mean Gaussian field with $[\Aout(t),\Aout^\dag(t')]=\delta(t-t')$, which factorizes into independent correlated sideband pairs $\pm \omega$~\cite{lvovsky2015squeezed} in the limit of long measurement time $T\to \infty$, with frequency modes spaced by $\Delta\omega=2\pi/T$~\cite{clerk2010introduction} [see Sec. II C of the Supplemental Material (SM)]. For each $\omega>0$, the pair is described by $\mathbf{r}=(A_{\text{out},\omega},A_{\text{out},-\omega},A^\dag_{\text{out},\omega},A^\dag_{\text{out},-\omega})^T$ and covariance matrix $\Sout=\mean{\mathbf{r}_i\mathbf{r}_j^\dag+\mathbf{r}_j^\dag \mathbf{r}_i}$, where
\fla{\boldsymbol{\Sigma}_{\text{out}}=(\mathbb{I}_4-\kappa \textbf{G}) \boldsymbol{\Sigma}_{\text{in}}(\mathbb{I}_4-\kappa \textbf{G})^\dag + \kappa \gamma \textbf{G} \boldsymbol{\Sigma}^{(B)}_{\text{in}} \textbf{G}^\dag,}
and $\Sin=\Sin^{(B)}=\mathbb{I}_4$ are the covariance matrices for the monitored input and the intrinsic loss channels, and $\mathbf{G}$ is the frequency-domain Green's function (see SM Sec. I). The smallest and largest eigenvalues of $\Sout$ define the output squeezing and antisqueezing spectra $\lambda_{s,a}(\omega)$, respectively. Let $\mathcal{I}_{\text{pair}}(\omega)$ denote the QFI carried by a correlated sideband pair. Independence of different frequency sectors makes the QFI additive, so that the stationary QFI rate at the monitored output is
\fla{\dot{\mathcal{I}}_{\text{out}}\equiv \lim_{T\to \infty} \frac{\mathcal{I}_{\text{out}}(T)}{T}=\int_0^\infty \frac{d\omega}{2\pi}\mathcal{I}_{\text{pair}}(\omega),\label{eq::Iout}}
and $\mathcal{I}_{\text{pair}}/2\pi$ defines the QFI density across the output spectrum, schematically shown in Fig.~\hyperref[fig1]{1(c)}. A balanced detuning-independent transformation of the two sidebands further factorizes each $\pm \omega$ sector into symmetric and antisymmetric Gaussian supermodes, defined by quadratures, $X_{\pm}=(X_{\text{out},\omega}\pm X_{\text{out},-\omega})/\sqrt{2},\; P_{\pm}=(P_{\text{out},\omega}\pm P_{\text{out},-\omega})/\sqrt{2},$
where $X_{\text{out},\pm\omega}=(A_{\text{out},\pm\omega}+A_{\text{out},\pm\omega}^\dag)/\sqrt{2}$ and $P_{\text{out},\pm\omega}=-i(A_{\text{out},\pm\omega}-A_{\text{out},\pm\omega}^\dag)/\sqrt{2}$ (see SM Sec. II B). The Wigner functions of the two supermodes $W(X_{\pm},P_{\pm})$ [inset of Fig.~\hyperref[fig1]{1(b)}] are identical up to a phase-space rotation. Figures~\hyperref[fig1]{1(b-c)} visualize this decomposition, where a perturbation in $\delta\omega$ deforms these Gaussian states, while $\Ipair$ resolves how this information is distributed in the emitted spectrum.

\textit{Asymptotic QFI scaling.} We now determine how the stationary output QFI scales and becomes distributed across the output spectrum as the parametric threshold is approached. Derivations of our results are given in the SM Sec. III.
For each correlated sideband pair, the Gaussian QFI can be directly evaluated using either the two-mode~\cite{vsafranek2015quantum} or single-mode formalism~\cite{serafini2023quantum}, yielding 
\fla{\mathcal{I}_{\text{pair}}(\omega)=\frac{16 \kappa  \xi ^2 \left[\Lambda_{\gamma}\left(\epsilon -\omega ^2\right)^2 +\kappa  \xi_c^2 \omega ^2 \Gamma^2\right]}{Q(\omega)^2 \left[2 \gamma  \kappa  \xi ^2+Q(\omega)\right]},}
where $\epsilon=\xi_c^2-\xi^2$ measures the distance from threshold, $\Lambda_\gamma=\delta \omega ^2 (2 \gamma +\kappa )+  \frac{\kappa\Gamma^2}{4}$, and $Q(\omega)=\omega ^2 \Gamma^2+(\epsilon -\omega ^2)^2$. We take the critical limit by increasing $\xi\to\xi_c$ at fixed $\kappa$, $\gamma$, and $\delta\omega$. Since $\xi_c>\delta\omega$ for finite $\Gamma$, for $\delta\omega \neq 0$ this path crosses the EP at $\xi=\delta\omega$ before reaching threshold. 
For a partially monitored OPO, i.e., with $\gamma>0$, the pair QFI at fixed $\omega\ne 0$ does not grow with $\nbar$, but the zero-frequency limit retains a quadratic photon-number dependence,
\fla{\mathcal{I}_{\text{pair}}(0)\sim \frac{32\Lambda_\gamma}{\gamma \xi_c^4}\bar{n}^2,\qquad \bar{n}\to \infty,}
where $\bar{n}=\xi^2/2\epsilon$ is the mean intracavity photon number. 
The same $\bar n^2$ scaling persists for $\omega\lesssim\epsilon/\Gamma$, so that the QFI enhancement is concentrated in a bandwidth $\Delta \omega_{\text{QFI}}\sim \epsilon/\Gamma \sim \xi_c^2/2\Gamma \bar{n}$, which shrinks as $1/\nbar$ [Fig.~\hyperref[fig1]{1(c)}]. Such spectral narrowing is a signature of the near-threshold critical enhancement. Consequently, integrating over the full stationary output spectrum gives
\fla{\dot{\mathcal{I}}_{\text{out}}\sim \frac{\Gamma \kappa+4\delta\omega^2}{\gamma\xi_c^2}\bar{n},\qquad \gamma>0.\label{eq::finite-rate}} 
The quadratic enhancement is therefore offset by the narrowing of the critically enhanced bandwidth, leaving the steady-state information rate linear in $\bar n$. Resolving the quadratic scaling around $\omega=0$ requires $\Delta\omega\lesssim\Delta\omega_{\text{QFI}}$, which itself shrinks with $\nbar$. Therefore, any fixed spectral resolution eventually fails to resolve this regime as threshold is approached. Replacing the discrete finite-$T$ modes by the continuum integral requires the stronger condition $\Delta\omega\ll\Delta\omega_{\text{QFI}}$. The same spectral narrowing appears in the antisqueezing spectrum, previously associated with critical slowing down of temporal fluctuations near a parametric threshold~\cite{gatti2017squeezing}. Here, the peak of the antisqueezing spectrum $\lambda_a(\omega)$ grows as $\bar n^2$ while its bandwidth shrinks as $1/\bar n$, mirroring the QFI spectrum (see SM Sec. IV A).

We now compare this output information with the ultimate constraint on the QFI imposed by inaccessible loss. General adaptive-metrology bounds~\cite{demkowicz2017adaptive}, as applied to monitored bosonic sensors in Refs.~\cite{alushi2024optimality,gorecki2025interplay,gualrecki2025time}, imply the following upper bound on the joint steady-state QFI of the cavity and monitored output $\mathcal{I}_{\rm cav+out}$:
\fla{\mathcal{I}_{\text{out}}(T)\le\mathcal{I}_{\text{cav+out}}(T)\le \frac{4}{\gamma} \int_0^T dt \; \langle a^\dag a \rangle _t = \frac{4 \bar{n} T}{\gamma}.\label{eq::bound}}
Here, the stationary protocol begins after the OPO has relaxed to steady state. The characteristic relaxation time is approximated by the slow eigenvalue of $\mathcal{E}$, which grows as $\tau_c\sim|\text{Re}\lambda_{+}|^{-1}\sim2\Gamma \nbar/\xi_c^2$ near threshold, so that a preparation stage lasting $\mathcal{O}(\tau_c)$ would contribute at most $\mathcal{O}(\bar n\tau_c/\gamma)$ under the same bound. A one-time preparation overhead becomes negligible since the continuum treatment requires $T\gg 2\pi/\Delta\omega_{\text{QFI}}\sim\tau_c$, such that $\mathcal{I}_{\text{prep}}/\mathcal{I}_{\text{out}}\to 0$. If preparation is included, the corresponding loose bound is $\mathcal{O}[\nbar (\tau_c+T)/\gamma]$. We define the protocol to begin after stationarity is established so that Eq.~\eqref{eq::bound} provides a strict constraint on the QFI accumulated specifically during the stationary measurement interval. The same steady-state bound applies to passive sensing strategies and coherently-seeded OPOs~\cite{gorecki2025interplay}, so at fixed $\nbar$, additional displacement cannot raise the leading information rate above $4\nbar/\gamma$. 

Crucially, the coefficient in Eq.~\eqref{eq::finite-rate} increases monotonically with $\delta\omega$,
and asymptotically approaches the fundamental value of $4/\gamma$ when $\delta\omega/\Gamma \gg 1 $. We focus on the physically relevant large-detuning limit at finite $\kappa,\gamma$, rather than the zero-loss limit $\Gamma\to0$, which simultaneously removes both intrinsic loss and the monitored output coupling. Large detuning does not eliminate the intracavity field, which remains a dynamical degree of freedom with near-threshold fluctuations that are correlated over a diverging time $\tau_c$. Likewise, for fixed nonzero intrinsic loss, the bound is approached for $\kappa/\gamma\gg1$, including in the far-detuned regime, provided the near-threshold condition $\epsilon\to0$ is maintained. Remarkably, in our stationary protocol, the monitored output alone asymptotically exhausts the joint cavity-monitored-output QFI bound in Eq.~\eqref{eq::bound}, capturing all leading-order metrological information that remains accessible after intrinsic loss. For comparison, consider a protocol based on direct measurements of the reduced stationary intracavity state, with QFI scaling as $\mathcal{I}_{\text{cav}}\sim \frac{4\delta\omega^2}{\Gamma\xi_c^2}\nbar \tau_c$ near threshold for any fixed $\delta\omega\ne 0$. After the cavity is prepared from vacuum over a relaxation time $\tau_c$, $\mathcal{I}_{\rm cav}$ does not continue to accumulate with $T$, whereas the stationary output record accumulates as $\mathcal{I}_{\text{out}}\sim \nbar T$. In principle, the same $\mathcal{O}(\nbar T)$ scaling could be reproduced from intracavity snapshots via repeated preparation and interrogation over $T/\tau_c$ cycles~\cite{alushi2024optimality}.
Here, a single long-time stationary output record approaches the bound. However, reaching the far-detuned or strongly overcoupled near-threshold regime at fixed $\gamma$ is increasingly demanding, since approaching threshold requires a correspondingly larger parametric drive $\xi$, set by the pump amplitude and $\chi^{(2)}$ nonlinearity in the illustrative optical realization. Moreover, writing the parametric drive as a fraction of threshold, e.g., $\xi^2=r\xi_c^2$, an asymptotic path must maintain $\epsilon=(1-r)\xi_c^2\to 0$, so that if $\xi_c$ grows, $r$ must approach unity more rapidly.

Interestingly, at fixed $\Gamma$, the EP and parametric threshold become asymptotically adjacent in the large-detuning limit, since $\xi_c \sim  \delta\omega+\frac{\Gamma^2}{8\delta\omega}$. These conditions, however, remain dynamically distinct for any finite $\Gamma$, and their asymptotic proximity does not make the enhancement EP-driven. Squeezing is likewise not a reliable indicator of metrological optimality. For arbitrary fixed $\kappa,\gamma,\delta\omega$, the minimum of $\lambda_s(\omega)$ converges to its floor of $\gamma/\Gamma$ on approaching threshold, with $1/\nbar^2$ corrections. The overcoupled regime saturating the QFI bound coincides with infinite squeezing $\gamma/\Gamma\to 0$, while the far-detuned regime saturates the QFI bound without changing the degree of asymptotic squeezing at fixed $\Gamma$.

In the fully monitored OPO limit, i.e., $\gamma=0$, the adaptive-metrology construction~\cite{demkowicz2017adaptive} no longer yields a bound linear in $T$ (see SM Sec. III B), so that faster-than-linear QFI accumulation with observation time is not prohibited. The asymptotic frequency-domain sideband pairs are then pure, and integrating their QFI over frequency gives 
\fla{\dot{\mathcal{I}}_{\text{out}}^{(\gamma=0)}\sim \frac{16 \kappa}{\xi_c^2}\bar{n}^3.}
The output-QFI rate scales cubically at exactly $\gamma=0$, but linearly for every fixed $\gamma>0$. The cubic rate is nevertheless not super-Heisenberg in the emitted-photon resource, since resolving the critically enhanced bandwidth requires $T\gtrsim 2\pi/\Delta \omega_{\text{QFI}}\sim\nbar$. With $N_{\text{out}}=\kappa\bar nT$,
\fla{\mathcal{I}_{\text{out}}=\begin{cases}
    \mathcal{O}(\bar{n} T)=\mathcal{O}(N_{\text{out}}),& \gamma> 0,\\
\mathcal{O}(\bar{n}^3 T)\lesssim\mathcal{O}(\bar{n}^2T^2)=\mathcal{O}(N_{\text{out}}^2),& \gamma=0.
\end{cases}}
Thus the fully monitored OPO does not exceed the Heisenberg scaling in the total number of photons emitted in the output channel. Our stationary continuum limit requires the stronger condition, $T\gg 2\pi/\Delta \omega_{\text{QFI}}$ or $T/\nbar\to \infty$, for which $\mathcal{I}_\text{out}/N^2_{\text{out}}\to 0$. Within our treatment, the OPO therefore does not access the QFI regime in which Heisenberg scaling could be saturated.

\begin{figure}[!t]
    \centering
    \includegraphics[width=1\linewidth]{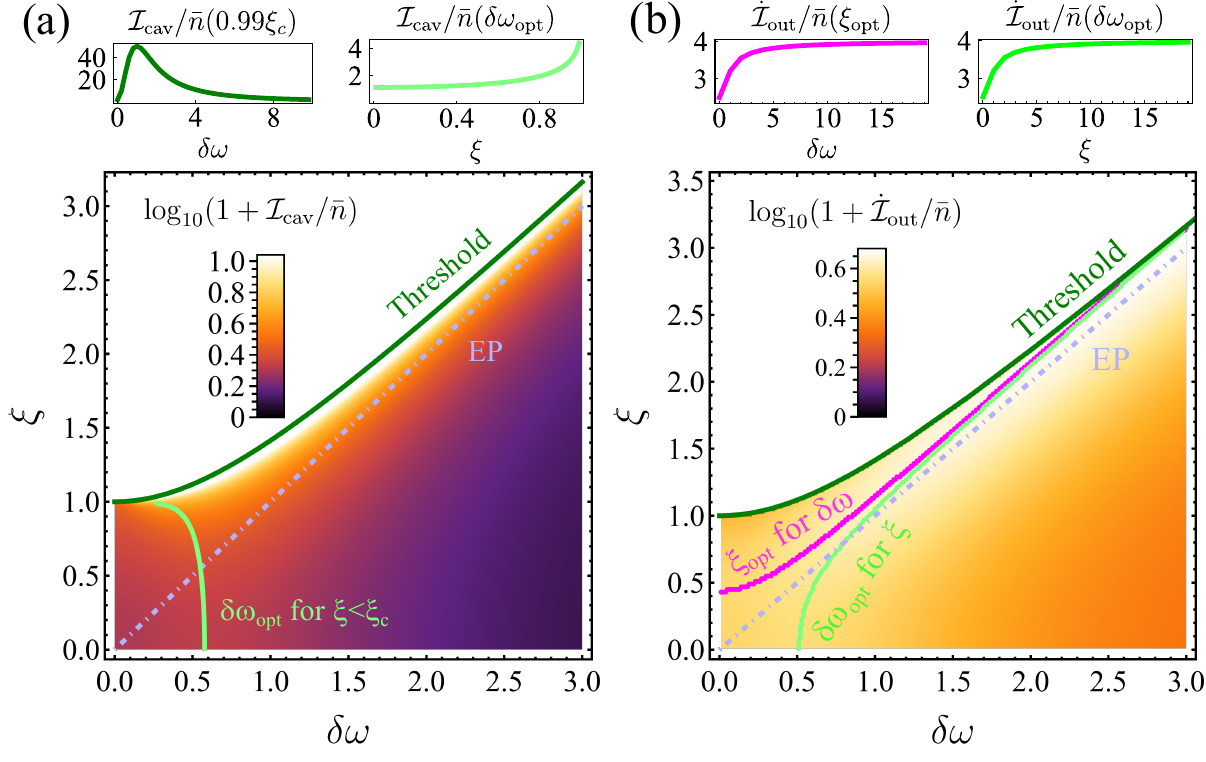}
    \vspace{-0.7cm}
    \caption{QFI across OPO parameter space. (a) Intracavity QFI per photon $\mathcal{I}_{\text{cav}}/\nbar$ and (b) output QFI rate per photon $\dot{\mathcal{I}}_{\text{out}}/\nbar$ versus $\delta\omega$ and $\xi$. In (a), the light green contour gives $\delta\omega_{\mathrm{opt}}$ for $\xi<1$; for $\xi>1$, the optimum for both $\xi$ and $\delta\omega$ lies arbitrarily close to threshold, so the top panels show $\mathcal I_{\mathrm{cav}}/\bar n$ at $\xi=0.99\xi_c$ and along $\delta\omega_{\mathrm{opt}}$. In (b), the magenta and light green contours give $\xi_{\text{opt}}$ and $\delta\omega_{\text{opt}}$; the top panels show the QFI rate per photon along these contours. Threshold is in solid dark green and the EP in dash-dotted purple. Density ranges are clipped for visibility. Fixed parameters: $\kappa,\gamma=1$.}
    \label{fig2}
\end{figure}

\textit{Per-photon QFI optimization.} Having established the asymptotic scaling, we turn to the information available per intracavity photon, separating sensing efficiency from the increase simply due to larger intracavity occupation. Our central result here is that neither the intracavity QFI nor the critical scaling alone can determine nonasymptotic operating points that maximize sensing efficiency. We compare $\mathcal{I}_{\text{cav}}/\nbar$, the information stored in the stationary intracavity state per stored photon, with $\dot{\mathcal{I}}_{\text{out}}/\nbar$, the output information rate per photon. First, we highlight that neither $\mathcal{I}_{\text{cav}}/\nbar$ [Fig.~\hyperref[fig2]{2(a)}] nor $\dot{\mathcal{I}}_{\text{out}}/\nbar$ [Fig.~\hyperref[fig2]{2(b)}] is maximized along the EP contour, so EP is generally a suboptimal operating point. A separate optimization over the outcoupling $\kappa$ likewise gives an optimum distinct from that maximizing squeezing or antisqueezing (see SM Sec. IV A).

At fixed nonzero $\delta\omega$, $\mathcal I_{\text{cav}}/\bar n$ increases continuously toward threshold [Fig.~\hyperref[fig2]{2(a)}], reflecting the $\bar n^2$ term in the intracavity QFI. The output behaves differently, and $\dot{\mathcal I}_{\text{out}}/\bar n$ is maximized at a finite $\xi_{\text{opt}}<\xi_c$, which moves progressively closer to threshold as $\delta\omega$ increases [Fig.~\hyperref[fig2]{2(b)}, top left]. Over the numerically accessible $\delta\omega$ range, the optimized rate per photon approaches the large-detuning bound of $4/\gamma$. By contrast, close to $\delta\omega=0$, the near-threshold asymptotic coefficient remains below the fundamental bound, so the optimum need not lie close to threshold. Further emphasizing the difference between
intracavity and output sensing, optimal $\delta\omega_{\text{opt}}$ for each $\xi$ differs for the intracavity and output QFI [Figs.~\hyperref[fig2]{2(a-b)}, top right]. Moreover, keeping a fixed fraction of threshold, such as $\xi=0.99\xi_c$, eventually loses the intracavity enhancement at large detuning [Fig.~\hyperref[fig2]{2(a)}, top left], since critical operation then requires tuning progressively closer to $\xi_c$.

\begin{figure}[!t]
    \centering
    \includegraphics[width=1\linewidth]{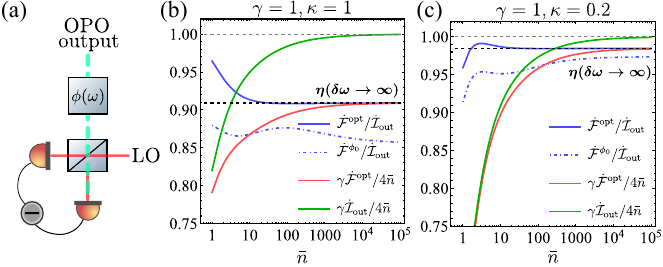}
    \vspace{-0.7cm}
    \caption{Homodyne detection maximizing CFI in a single record. (a)~A prescribed frequency-dependent phase shift applied to the OPO output realizes the optimal quadrature angle at each $\omega$ relative to a fixed-phase, single-frequency coherent LO at $\omega=0$. (b,c) QFI and CFI along the near-threshold far-detuned path with $\delta\omega=\Gamma\nbar^{1/4}$ and $\xi=\xi_c\sqrt{2\nbar/(2\nbar+1)}$. Solid and dash-dotted blue curves show the extraction efficiencies for homodyne measurements with and without an optimized frequency-dependent shift, respectively. The latter case's constant phase is set to the optimal phase $\phi_0$ for $\omega=0$; red and green curves show the corresponding CFI and QFI rates normalized to the bound \(4\bar n/\gamma\). Horizontal dashed lines indicate the predicted asymptotic values.}
    \label{fig3}
\end{figure}

\textit{Single-record homodyne readout to approach QFI.} 
Lastly, we ask how closely the output QFI can be approached in a single record using a simple balanced-homodyne receiver driven by a single-frequency coherent local oscillator (LO) at the rotating-frame reference frequency [Fig.~\hyperref[fig3]{3(a)}]. By a “single record,” we mean that all frequency bins are read out from the Fourier components of one long-time photocurrent trace acquired with a fixed receiver configuration. As illustrated in Fig.~\hyperref[fig3]{3(a)}, the prescribed quadrature angle at each analysis frequency is realized by applying a spectral phase shift to the broadband output before mixing it with the monochromatic LO, requiring neither phase scanning nor measurement reconfiguration during acquisition.
For $\gamma=0$, each supermode is pure, and homodyne detection is known to attain the QFI for pure Gaussian covariance models with fixed first moments~\cite{monras2013phase}. 
For $\gamma>0$, purity itself can vary, so homodyne detection need not capture the full covariance response. The optimal phase space directions at each $\omega$ follow from the generalized eigenproblem  $\partial_{\delta\omega}[\boldsymbol{\Sigma}_{\pm}(\omega)]\mathbf{u}=\alpha_{1,2}\boldsymbol{\Sigma}_{\pm}(\omega)\mathbf{u}$, where $\boldsymbol{\Sigma}_{\pm}(\omega)$ is the covariance matrix of either supermode. The classical Fisher information (CFI) is then $\mathcal{F}_{\rm pair}^{\rm opt}(\omega)=\max (\alpha_1^2,\alpha_2^2)$ for each $\pm\omega$.

Near-threshold, the fraction of output QFI extracted by phase-optimized homodyne detection, $\eta=\dot{\mathcal{F}}^{\rm opt}/\dot{\mathcal{I}}_{\rm out}$, with $\dot{\mathcal{F}}^{\rm opt}$ defined analogously to Eq.~\eqref{eq::Iout}, approaches
\fla{\eta\sim \frac{1}{2}+\frac{1}{\pi} \arctan \left(\frac{\delta\omega/\xi_c}{\sqrt{\kappa/\gamma}}\right)+\frac{(\delta\omega/\xi_c)\sqrt{\kappa/\gamma}}{\pi (\delta\omega^2/\xi_c^2+\kappa/\gamma)},}
and $\eta$ increases monotonically with $\frac{\delta\omega/\xi_c}{\sqrt{\kappa/\gamma}}$. Such parametric dependence of $\eta$ distinguishes the two limits in which the output QFI reaches the fundamental bound. For $\kappa/\gamma\gg 1$, the constraint $\delta\omega/\xi_c\le 1$ forces the asymptotic extraction efficiency to approach $1/2$. The covariance response becomes dominated by a rotation of the noise ellipse, and the two generalized eigenvalues satisfy \(\alpha_1\simeq-\alpha_2\). For a highly mixed output, both contribute to the QFI, $\mathcal{I}_{\rm pair}(\omega)\sim \alpha_1^2+\alpha_2^2$,  whereas a single optimized homodyne quadrature can only access one of these directions. The resulting $50\%$ efficiency is consistent with phase estimation of highly mixed squeezed thermal states, for which homodyne asymptotically extracts half of the QFI~\cite{oh2019optimal}. By contrast, large detuning drives $\eta\to\frac{1}{2}+\frac{1}{\pi}\arctan\sqrt{\frac{\gamma}{\kappa}}+\frac{\sqrt{\kappa/\gamma}}{\pi (1+\kappa/\gamma)}$, which approaches unity with $\kappa/\gamma\ll 1$. Hence, weak outcoupling at fixed $\gamma$ can make our detection nearly optimal in the far-detuned regime, while the output QFI rate remains close to the same fundamental bound $4\nbar/\gamma$. At the same time, decreasing $\kappa$ 
should still satisfy the underlying assumptions of the asymptotic QFI treatment (see SM Sec. III, IV C). 

Figures~\hyperref[fig3]{3(b-c)} show this convergence along a near-threshold, far-detuned path with $\delta\omega=\Gamma \nbar^{1/4}$. For $\kappa=\gamma=1$, as in Fig.~\ref{fig2}, the asymptotic phase-optimized homodyne efficiency (blue curve) is $\sim91\%$, and reducing $\kappa$ raises it to $\sim98\%$ [Fig.~\hyperref[fig3]{3(c)}] without reducing the asymptotically attainable QFI rate. A homodyne without the prescribed frequency-dependent phase shift in the OPO output remains generally suboptimal (blue dash-dotted curve). However, when $\frac{\kappa/\gamma}{\delta\omega^2/\xi_c^2}\ll1$, the optimal quadrature directions have the same leading-order orientation in the principal-axis basis throughout the critical bandwidth. Along the large-detuning path considered here, the principal-axis orientations at different frequencies within the critical bandwidth also agree to leading order. Consequently, for $\kappa/\gamma \ll 1$, the frequency dependence of the optimal detection quadratures across the critical bandwidth is parametrically small, allowing the case without the spectral phase shaping to remain nearly optimal along the asymptotic path (see SM Sec. IV C).

\textit{Discussion.}
Taken together, our results establish the stationary OPO output as a multimode metrological resource. In a partially monitored OPO near threshold, individual sideband frequency pairs carry QFI scaling as \(\bar n^2\) within a bandwidth shrinking as \(1/\bar n\), yielding a linear integrated information rate. The same narrowing appears in the antisqueezing spectrum, providing an experimentally visible signature of the critical enhancement. In the far-detuned or strongly overcoupled regimes, the monitored output asymptotically reaches the joint cavity-output QFI bound to leading order. The same \(\mathcal O(\bar nT)\) bound applies to much more general adaptive strategies under inaccessible Markovian loss~\cite{demkowicz2017adaptive}, and can also be saturated by passive coherent-state sensing under appropriate conditions (see SM Sec. III C). Here, the OPO reaches the same leading information rate without a coherent probe, entirely through quantum fluctuations and correlated sidebands. Our results identify that critical dynamics of dissipative parametric sensors can both concentrate the available metrological information in the accessible output and, in favorable regimes, make it directly retrievable with homodyne detection. 

\begin{acknowledgments}

We thank Nicol\'as Quesada, Zhongan Lin, and Evgeny Moiseev for useful discussions. We acknowledge support from Qu\'ebec’s Minist\`ere de l’\'Economie, de l’Innovation et de l’\'Energie (MEIE) and the Natural Sciences and Engineering Research Council of Canada (NSERC) under RGPIN-2023-03630 and ALLRP 588334-23. This research was undertaken, in part, thanks to funding from the Canada Research Chairs Program (CRC-2024-00338). P. B. acknowledges support from the Fonds de recherche du Québec – Nature et technologies (FRQNT) through a Master’s Research Scholarship. This work benefited from the research communities fostered by the Institut transdisciplinaire d’information quantique (INTRIQ) and the Regroupement québécois sur les matériaux de pointe (RQMP), both part of the FRQNT Regroupements stratégiques program.
 Generative AI tools were used to assist with language editing, literature searches, code refinement, and checks of analytical derivations. All AI-assisted material was independently reviewed and verified by the authors, who take full responsibility for the content and integrity of this manuscript.
\end{acknowledgments}

\bibliographystyle{style}

\bibliography{references_abbreviated}

\end{document}



\title{Supplemental Material for "Critical quantum metrology in the output of an optical parametric oscillator"}

\date{\today}

\maketitle
\tableofcontents

\section{Output Covariance Matrix of a Single-Mode Optical Parametric Oscillator}\label{sec:outA}
\subsection{Fully Monitored, $\gamma=0$}\label{sec:full-monitor}
We first consider an optical parametric oscillator (OPO) where the only dissipation arises from coupling to the input-output port, i.e., there is no intrinsic loss ($\gamma=0$ in the main text). The lab frame Hamiltonian is
\fla{\mathcal{H}=\omega_c a^\dag a+ \frac{i\xi}{2}\left((a^\dag)^2e^{-i\omega_p t} - a^2 e^{i\omega_p t} \right),}
where $\omega_{c,p}$ is the cavity or pump frequency, respectively, $\xi$ is the parametric drive strength, chosen real and positive, and $\hbar=1$. The Langevin equation of motion is 
\fla{\dot{a}=-i\omega_c a+\xi a^\dag e^{-i\omega_p t}-\frac{\kappa}{2}a+\sqrt{\kappa} A_{\rm in},}
where $\kappa$ is the decay rate through the monitored input-output port. We work in the rotating frame $a\to e^{i\omega_p t/2}a$, so that the resulting dynamical equations are
\fla{\begin{pmatrix}
\dot{a} \\ \dot{a}^\dag 
\end{pmatrix}
=\begin{pmatrix}
    -\frac{\kappa}{2} -i\delta \omega & \xi \\ \xi & -\frac{\kappa}{2}+ i\delta \omega 
\end{pmatrix} \begin{pmatrix}
    a \\ a^\dag 
\end{pmatrix} +\sqrt{\kappa} \begin{pmatrix}
    A_{\rm in} \\ A_{\rm in}^\dag 
\end{pmatrix} , \label{eq::langevin}}
where $\delta \omega=\omega_c-\frac{\omega_p}{2}$ is the cavity detuning from half the pump frequency, and we assume $\delta\omega \ge 0$. We work in the stable stationary regime. The output is a continuous traveling field, which we analyze in the frequency domain. We adopt the following Fourier transform convention 
\fla{\begin{pmatrix}
    \mathcal{F}[a] \\ \mathcal{F}[a^\dag]
\end{pmatrix}=\begin{pmatrix}
    a(\omega) \\ a^\dag (\omega)
\end{pmatrix} = \begin{pmatrix}
    \int e^{i\omega t } a(t)dt \\ \int e^{i\omega t} a^\dag (t)dt
\end{pmatrix},}
such that $[a(\omega)]^\dag = a^\dag (-\omega)$ and $a(t)=\frac{1}{2\pi} \int e^{-i\omega t} a(\omega) d\omega $. Note that this is not the usual quantum optics convention where one has $[a(\omega)]^\dag=a^\dag(\omega)$, which is defined with opposite phases for the transforms of $a$ and $a^\dag$. Fourier transforming the Langevin equation for $\dot{a}$, with $\int dt \; e^{i\omega t} \dot{a}=-i\omega a(\omega)$, gives
\fla{-i\omega a(\omega) = \left( -\frac{\kappa}{2} - i\delta \omega \right)a(\omega) + \xi a^\dag (\omega) +\sqrt{\kappa} A_{\rm in}(\omega).} 
For fixed $\omega$, $a(\omega)$ and $a^\dag(-\omega)=[a(\omega)]^\dag$ form a creation-annihilation pair associated with that $\omega$, while the OPO dynamics couple opposite sidebands, $a(\omega)$ and  $a^\dag(\omega)$. The resulting system of equations is 
\fla{
\begin{pmatrix}
    a(\omega) \\ a(-\omega) \\ a^\dag(-\omega) \\ a^\dag(\omega)
\end{pmatrix} = \sqrt{\kappa} \underbrace{\begin{pmatrix}
    -i\omega +i\delta \omega +\frac{\kappa}{2} & 0 & 0 & -\xi \\
    0 & i\omega+i\delta \omega + \frac{\kappa}{2}&-\xi&0\\
    0 & -\xi& i\omega - i\delta \omega + \frac{\kappa}{2} & 0 \\
    -\xi & 0 & 0 & -i\omega -i\delta \omega + \frac{\kappa}{2}
\end{pmatrix}^{-1}}_{\mathbf{G}_A}\begin{pmatrix}
    A_{\rm in}(\omega) \\ A_{\rm in}(-\omega) \\ A_{\rm in}^\dag(-\omega) \\ A_{\rm in}^\dag(\omega)
\end{pmatrix},
}
where $\mathbf{G}_A$ is $\mathbf{G}$ in the main text. We then define the input, output, and intracavity operator vectors for $\omega>0$ as 
\fla{\mathbf{A}_{\rm in}&=\left(A_{\rm in}(\omega),A_{\rm in}(-\omega),A_{\rm in}^\dag(-\omega),A_{\rm in}^\dag(\omega) \right)^T=\left(\Ain,\Atin,\Ain^\dag,\Atin^\dag\right)^T,\\
\mathbf{A}_{\rm out}&=\left(A_{\rm out}(\omega),A_{\rm out}(-\omega),A_{\rm out}^\dag(-\omega),A_{\rm out}^\dag(\omega) \right)^T=\left(\Aout,\Atout,\Aout^\dag,\Atout^\dag\right)^T, \\
\mathbf{A}&=\left(a(\omega),a(-\omega),a^\dag(-\omega),a^\dag(\omega) \right)^T,}
where we labeled $A_{1,\rm out}=A_{\rm out}(\omega),\; A_{2,\rm out}=A_{\rm out}(-\omega),\; A_{1,\rm out}^\dag=A_{\rm out}^\dag(-\omega),\; A_{2,\rm out}^\dag=A_{\rm out}^\dag(\omega)$, and analogously for the inputs, so that the two-mode picture is explicit. The input-output relation is $\mathbf{A}_{\rm out}=\mathbf{A}_{\rm in}-\sqrt{\kappa} \mathbf{A}=\left( \mathbb{1}-\kappa \mathbf{G}_A\right)\mathbf{A}_{\rm in}$. Note that in a fully monitored system, the transformation between input and output is symplectic, such that $(\mathbb{1}-\kappa \textbf{G}_A)\textbf{K}(\mathbb{1}-\kappa \textbf{G}_A)^\dag=\textbf{K}$ with $\textbf{K}=\operatorname{diag}(1,1,-1,-1)$ the commutation metric. Hence, for a vacuum input, each joint nonzero $\pm\omega$ output sector is a pure two-mode Gaussian state. The output covariance matrix is defined as
\fla{\SoutA&=\mean{\mathbf{A}_{\rm out}^i (\mathbf{A}_{\rm out}^j)^\dag+(\mathbf{A}_{\rm out}^j)^\dag \mathbf{A}_{\rm out}^i}.\label{eq::out-sigma}}
Note the absence of $\frac{1}{2}$ in our definition of covariance. In our notation, we suppress the delta-function correlations and interpret $\boldsymbol{\Sigma}(\omega)$ as the covariance matrix at the frequency of interest; \ref{sec:discrete} derives this representation from normalized finite-time modes. Equation~\eqref{eq::out-sigma} can be expressed in terms of the input covariance 
\fla{\SoutA
&=\left(\mathbb{1}-\kappa \mathbf{G}_A\right)\boldsymbol{\Sigma}_{\rm in,A} \left( \mathbb{1}-\kappa \mathbf{G}_A\right)^\dag=\left( \mathbb{1}-\kappa \mathbf{G}_A\right)\left( \mathbb{1}-\kappa \mathbf{G}_A\right)^\dag,
}
where $\boldsymbol{\Sigma}_{\rm in,A}=\mathbb{1}$ is the input covariance matrix for the two-mode vacuum in the creation/annihilation operator basis. We next express the same transformation in the quadrature basis. We consider operators 
\fla{x(\omega)=\frac{1}{\sqrt{2}}\left( a(\omega) + a^\dag (-\omega)\right),\qquad
p(\omega)=-\frac{i}{\sqrt{2}}\left( a(\omega) - a^\dag (-\omega)\right),}
which are Hermitian quadratures of the resolved-frequency modes, rather than Fourier transforms of the time-domain quadrature operators. Moreover, $x(\omega) = [x(\omega)]^\dag$ and $p(\omega)=[p(\omega)]^\dag$, and similarly for input/output quadratures.
The resulting system of equations is
\fla{\begin{pmatrix}
    x(\omega) \\x(-\omega) \\ p(\omega) \\ p(-\omega)
\end{pmatrix} = \sqrt{\kappa}\underbrace{\begin{pmatrix}
    \frac{\kappa}{2} & -\xi & \omega - \delta \omega & 0 \\ -\xi & \frac{\kappa}{2} & 0 & -\omega - \delta \omega \\
    \delta \omega - \omega & 0 & \frac{\kappa}{2} & \xi \\ 0 & \delta \omega + \omega & \xi & \frac{\kappa}{2}
\end{pmatrix}^{-1}}_{\textbf{G}_R} \begin{pmatrix}
    X_{\rm in}(\omega) \\ X_{\rm in}(-\omega) \\ P_{\rm in}(\omega) \\ P_{\rm in}(-\omega) 
\end{pmatrix}.}
Defining
$
\mathbf{R}_{\mathrm{in/out}}=
\left(
X_{1,\mathrm{in/out}},
X_{2,\mathrm{in/out}},
P_{1,\mathrm{in/out}},
P_{2,\mathrm{in/out}}
\right)^T,
$ with $X_{1,\rm in/out}=X_{\rm in/out}(\omega)$, $X_{2,\rm in/out}=X_{\rm in/out}(-\omega)$, and similarly for $P_{1,2}$, the input-output relation is $
\mathbf{R}_{\mathrm{out}}
=
\left(\mathbb{1}-\kappa \mathbf{G}_R\right)\mathbf{R}_{\mathrm{in}}$. To transform to the creation/annihilation basis, we first relate the operators as
\fla{\textbf{R}=\begin{pmatrix}
    x(\omega) \\ x(-\omega) \\ p(\omega) \\ p(-\omega) 
\end{pmatrix}& = \underbrace{\frac{1}{\sqrt{2}}\begin{pmatrix}
    1& 0 & 1 & 0\\
    0 & 1 & 0 & 1 \\
    -i & 0 & i & 0\\
    0 & -i & 0 & i
\end{pmatrix}}_{\mathbf{U}}\begin{pmatrix}
    a(\omega) \\ a(-\omega) \\ a^\dag (-\omega) \\ a^\dag (\omega) 
\end{pmatrix}=\mathbf{U}\mathbf{A}
\label{basis}.}
Using the transformation $\mathbf U$, one obtains $\textbf{G}_A=\textbf{U}^\dag \textbf{G}_R \textbf{U}$ and the relation $(\mathbb{1}-\kappa \textbf{G}_A)=\textbf{U}^\dag (\mathbb{1}-\kappa \textbf{G}_R)\textbf{U}$. The two-mode symmetrized covariance matrices are then related as
\fla{\Sout=(\mathbb{1}-\kappa \textbf{G}_R)(\mathbb{1}-\kappa \textbf{G}_R)^T,\qquad
\SoutA=\textbf{U}^\dag \Sout \mathbf{U}=(\mathbb{1}-\kappa \textbf{G}_A)(\mathbb{1}-\kappa \textbf{G}_A)^\dag.}

\subsection{Partially Monitored, $\gamma\ne 0$}\label{sec:outLoss}
We now include intrinsic loss through an independent, unmonitored bath $B_{\rm in}$ coupled at rate $\gamma$. The Langevin equations that account for intrinsic loss are 
\fla{\begin{pmatrix}
\dot{a} \\ \dot{a}^\dag 
\end{pmatrix}
=\underbrace{\begin{pmatrix}
    -\frac{\Gamma}{2} -i\delta \omega & \xi \\ \xi & -\frac{\Gamma}{2}+ i\delta \omega 
\end{pmatrix}}_{\boldsymbol{\mathcal{E}}} \begin{pmatrix}
    a \\ a^\dag 
\end{pmatrix} +\sqrt{\kappa} \begin{pmatrix}
    A_{\rm in} \\ A_{\rm in}^\dag 
\end{pmatrix} +\sqrt{\gamma} \begin{pmatrix}
    B_{\rm in} \\ B_{\rm in}^\dag 
\end{pmatrix}  \label{eq::intrinsic},}
where $\Gamma=\kappa+\gamma$ and $\boldsymbol{\mathcal{E}}$ is the dynamical matrix.
The resulting input-output relations are 
\fla{\mathbf{A}&=\textbf{G}_A \left[ \sqrt{\kappa}\mathbf{A}_{\rm in}+\sqrt{\gamma} \textbf{B}_{\rm in}\right],\\
\mathbf{A}_{\rm out}&=(\mathbb{1}-\kappa \textbf{G}_A) \mathbf{A}_{\rm in} - \sqrt{\kappa \gamma} \textbf{G}_A \textbf{B}_{\rm in},}
where $\mathbf{B}_{\rm in}=\left(B_{\rm in}(\omega),B_{\rm in}(-\omega),B_{\rm in}^\dag(-\omega),B_{\rm in}^\dag(\omega) \right)^T$. Here, $\mathbf G_A$ is the Green's function matrix defined as in~\ref{sec:full-monitor}, but with $\kappa$ replaced by $\Gamma$. The two baths are uncorrelated, so the output covariance becomes
\fla{\SoutA=(\mathbb{1}-\kappa \textbf{G}_A) \boldsymbol{\Sigma}_{\rm in,A}(\mathbb{1}-\kappa \textbf{G}_A)^\dag + \kappa \gamma \textbf{G}_A \boldsymbol{\Sigma}^{(B)}_{\rm in,A} \textbf{G}_A^\dag,}
where $\boldsymbol{\Sigma}_{\rm in,A}=\boldsymbol{\Sigma}^{(B)}_{\rm in,A}=\mathbb{1}_4$ for vacuum covariance matrices of both loss channels. Furthermore, $\langle\mathbf A_{\rm in}\rangle=\langle\mathbf B_{\rm in}\rangle=0$, and hence $\langle\mathbf a\rangle=\langle\mathbf A_{\rm out}\rangle=0$. The quantum Fisher information (QFI) therefore depends only on second moments.

\section{Quantum Fisher Information}\label{sec:QFI}
\subsection{Two-Mode Representation}
Each correlated nonzero $\pm\omega$ sector of the output covariance matrix is a two-mode Gaussian state. For a mixed output state, i.e., $\gamma>0$, the zero-mean two-mode Gaussian QFI with respect to $\delta\omega$ is~\cite{vsafranek2015quantum}
\fla{\mathcal{I}&=\frac{1}{2(|\mathbf{A}|-1)}\left(|\mathbf{A}| \operatorname{Tr} \left[(\mathbf{A}^{-1} \mathbf{A}')^2\right] +\sqrt{|\mathbb{1}+\mathbf{A}^2|} \operatorname{Tr} \left[((\mathbb{1}+\mathbf{A}^2)^{-1} \mathbf{A}')^2\right]+4(\nu_1^2-\nu_2^2)\left( -\frac{\nu_1'^2}{\nu_1^4-1}+\frac{\nu_2'^2}{\nu_2^4-1}\right)\right),}
where $\mathbf{A}=\textbf{K}\boldsymbol{\Sigma}_{\rm out,A}$, $\textbf{K}=\operatorname{diag}(1,1,-1,-1)$, primes denote derivatives with respect to the perturbed parameter $\delta\omega$, and $|\cdot|$ is the determinant. The symplectic eigenvalues of $\boldsymbol{\Sigma}_A$ are
\fla{\nu_{1,2}=\frac{1}{2} \sqrt{\operatorname{Tr}[\mathbf{A}^2]\pm \sqrt{(\operatorname{Tr}[\mathbf{A}^2])^2-16|\mathbf{A}|}}.}
For $\gamma=0$, the output state remains pure under variation of $\delta\omega$, and the QFI reduces to
\fla{
\mathcal{I}&=\frac{1}{4}\operatorname{Tr}[(\boldsymbol{\Sigma}_{\rm out,A}^{-1}\boldsymbol{\Sigma}'_{\rm out, A})^2] =\frac{1}{4}\operatorname{Tr}[(\boldsymbol{\Sigma}_{\rm out,R}^{-1}\boldsymbol{\Sigma}'_{\rm out,R})^2].
}
Within the stable region, the QFI is a smooth function of system
parameters, with the $\omega=0$ sector diverging as the parametric threshold is approached. Threshold occurs when a pole of the Green function reaches the real-frequency axis, which occurs at the value of the parametric drive $\xi_c^2=\delta\omega^2+\Gamma^2/4$.

\subsection{Single-Mode Representation}\label{SMR}
The supermode basis defined by collective quadratures $X_\pm=\frac{\Xout \pm \Xtout}{\sqrt{2}}$ and $P_\pm=\frac{\Pout\pm \Ptout}{\sqrt{2}}$ is obtained by $\textbf{S}\Sout \textbf{S}^T=\Sp\oplus\Sm$ with $\textbf{S}=\frac{1}{\sqrt{2}}\left(
\begin{array}{cccc}
 1 & 1 & 0 & 0 \\
 0 & 0 & 1 & 1 \\
 1 & -1 & 0 & 0 \\
 0 & 0 & 1 & -1 \\
\end{array}
\right)$. The two blocks satisfy $\boldsymbol{\Sigma}_-=\mathbf J\boldsymbol{\Sigma}_+\mathbf J^T$, with $\mathbf J=\begin{pmatrix}0&-1\\1&0\end{pmatrix}$, and are therefore related by a parameter-independent phase-space rotation. Hence they have equal purity and equal QFI. This transformation is symplectic (after a basis reordering), such that the new density matrix is related to the old one via a unitary transformation~\cite{adesso2014continuous}, which leaves the QFI invariant. The QFI of each supermode can be calculated with single-mode Gaussian state QFI~\cite{serafini2023quantum}
\fla{\mathcal{I}_{\pm}=\frac{1}{2} \frac{\operatorname{Tr}[(\boldsymbol{\Sigma}_\pm^{-1}\boldsymbol{\Sigma}'_\pm)^2]}{1+\mu^2}+\frac{2 \mu'^2}{1-\mu^4},\label{eq::sm-qfi}}
where $\mu=1/\sqrt{\det(\boldsymbol{\Sigma}_{\pm})}$ is the purity. The block-diagonal Gaussian covariance implies a product state of the two supermodes, so QFI additivity gives $\Ipair=\mathcal{I}_{+}+\mathcal{I}_{-}=2\mathcal{I}_{+}$, where $\Ipair$ is the two-mode QFI of $\pm \omega$ state. When the state is pure, i.e., $\gamma=0$, the single-mode QFI is instead $\mathcal{I}_{\pm}=\frac{1}{4} \operatorname{Tr}[(\boldsymbol{\Sigma}_\pm^{-1}\boldsymbol{\Sigma}'_\pm)^2]$. At $\omega=0$, the two sidebands coincide and there is only one physical frequency mode. Its independently calculated $2\times2$ covariance equals $\Sp(0)$. Hence the single-mode quantities derived from $\Sp$ apply directly at $\omega=0$ with $X_+\longleftrightarrow X$ and $P_+\longleftrightarrow P$ for $X=\frac{A_{\rm out}(0)+A_{\rm out}^\dag(0)}{\sqrt{2}}$ and $P=-i\frac{A_{\rm out}(0)-A_{\rm out}^\dag(0)}{\sqrt{2}}$. Therefore, the QFI of the physical $\omega=0$ mode is $\mathcal{I}_0=\mathcal{I}_+(0)=\frac{1}{2}\Ipair(0)$.

\subsection{Integrated QFI at the Output}\label{sec:discrete}
The total output QFI receives contributions from the full stationary spectrum. For a measurement record of duration $T$, we introduce normalized Fourier-bin modes~\cite{lvovsky2015squeezed,clerk2010introduction} 
\fla{A_n=\frac{1}{\sqrt{ T}}\int_{t_j}^{t_j+T} dt\; e^{i\omega_n t} A_{\rm out}(t),}
which form an orthonormal basis on the interval and satisfy
$[A_n,A_m^\dag]=\frac{1}{T} \int_{t_j}^{t_j+T} dt \; e^{i(\omega_n-\omega_m) t}=\delta_{nm}
$
for $\omega_n=2\pi n/T$ and spacing $\Delta\omega=2\pi/T$. The $n=0$ bin is therefore also a normalized single bosonic mode, but unlike $n>0$, it is not paired with a distinct $-n$ mode. In frequency space, using $A_{\rm out}(t)=\frac{1}{2\pi}\int d\omega\; A_{\rm out}(\omega)e^{-i\omega t}$, we equivalently have
$A_n=\int \; \frac{d\omega}{2\pi} \tilde{\phi}_n(\omega)A_{\rm out}(\omega)$, where
$\tilde{\phi}_n(\omega)=\sqrt{T} e^{i(\omega_n-\omega)(T/2+t_j)} \operatorname{sinc}\left[(\omega_n-\omega)T/2\right]$. In our framework, the OPO is allowed to reach its stationary state, and then a record of duration $T$ is collected.
In the limit of $T\to\infty$, 
\fla{\tilde{\phi}_n^*(\omega)\tilde{\phi}_n(\omega)=T \text{ sinc}^2[(\omega_n-\omega)T/2]\rightarrow 2\pi \delta (\omega_n-\omega).}
For $n>0$, defining $\mathbf{A}_n=\left(A_{\rm out,n},A_{\rm out,-n},A_{\rm out,n}^\dag,A_{\rm out,-n}^\dag\right)^T$, the discrete covariance $\boldsymbol{\Sigma}_n$ approaches $\SoutA(\omega_n)$ as $T\to\infty$. For finite $T$, each Fourier bin samples a finite frequency window centered around $\omega_n$. As $T\to\infty$, the stationary OPO correlations occur only between opposite-frequency pairs. The covariance therefore becomes block diagonal in the $\pm\omega_n$ sectors, and the QFI is additive over $n>0$
\fla{\sum_{n>0} \Ipair(\omega_n)\to\frac{1}{\Delta \omega} \int_0^\infty d\omega \; \Ipair(\omega)=\frac{T}{2\pi} \int_0^\infty d\omega \; \Ipair(\omega).}
The output QFI rate in steady-state considered in the main text is 
\fla{\dot{\mathcal{I}}_{\rm out}=\lim_{T\to \infty} \left(\frac{\mathcal{I}_0+\sum_{n>0}\Ipair(\omega_n)}{T}\right)=\frac{1}{2\pi}\int_0^\infty d\omega \; \Ipair(\omega).
}
Here, $\mathcal{I}_0$ is the single-mode QFI of the $n=0$ bin discussed in \ref{SMR}. Below threshold, $\mathcal{I}_0$ remains finite as $T\to\infty$, so that $\mathcal{I}_0=\mathcal{O}(1)$, and the first term can be neglected. Later, when we take the limit of $\nbar \to \infty$, we do so sequentially, i.e., the $T\to \infty$ limit is taken first.

\section{Quantum Fisher 
Information Scaling with Mean Photon Number}\label{sec:scaling}

\subsection{Intracavity QFI}
The intracavity covariance matrix $\boldsymbol{\Sigma}$, defined analogously to $\boldsymbol{\Sigma}_{\rm out,R}$, evolves according to the diffusion equation~\cite{serafini2023quantum}
\fla{\dot{\boldsymbol{\Sigma}} = \boldsymbol{\mathcal{E}} \boldsymbol{\Sigma} + \boldsymbol{\Sigma} \boldsymbol{\mathcal{E}}^T+\boldsymbol{\mathcal{D}},}
where $\boldsymbol{\mathcal{D}}=\Gamma \mathbb{1}_2$ and $\boldsymbol{\mathcal{E}}=\left(
\begin{array}{cc}
 \xi -\frac{\Gamma}{2} & \delta \omega  \\
 -\delta \omega  & -\frac{\Gamma}{2} -\xi  \\
\end{array}
\right)$ is the dynamical matrix in quadrature basis $\mathbf{R}=(x,p)^T$. We use the same vacuum-normalized covariance convention as above with $\boldsymbol{\Sigma}_{mn}=\mean{\left\{ \mathbf{ R}_m, \mathbf{R}_n\right\}}$. In the stable regime, i.e., $\xi<\xi_c$, the steady-state intracavity covariance matrix is obtained from solving $\dot{\boldsymbol{\Sigma}}=0$, which is explicitly
\fla{\boldsymbol{\Sigma}(t\to \infty)=\left(
\begin{array}{cc}
 \frac{\Gamma (\Gamma +2 \xi )+4 \delta \omega ^2}{\Gamma^2+4 \left(\delta\omega^2-\xi^2\right)} & -\frac{4 \delta \omega  \xi }{\Gamma^2+4 \left(\delta\omega^2-\xi^2\right)} \\
 -\frac{4 \delta \omega  \xi }{\Gamma^2+4 \left(\delta\omega^2-\xi^2\right)} & \frac{\Gamma (\Gamma -2 \xi )+4 \delta \omega ^2}{\Gamma^2+4 \left(\delta\omega^2-\xi^2\right)} \\
\end{array}
\right).}
The steady-state intracavity QFI is obtained by applying the single-mode Gaussian QFI formula~\eqref{eq::sm-qfi}
\fla{\mathcal{I}_{\rm cav}=\frac{8 \xi^2 \left(\Gamma ^2-4 \xi^2+12 \delta \omega^2\right)}{\left(\Gamma ^2-4 \xi^2+4 \delta \omega^2\right)^2 \left(\Gamma ^2-2 \xi^2+4 \delta \omega^2\right)}=\frac{1}{\xi_c^2-\xi^2/2} \left( \nbar+4 \frac{\delta \omega^2}{\xi^2} \nbar^2\right),}
where $\nbar$ is the mean intracavity photon number in steady-state, found from
\fla{\nbar&=\frac{1}{4}\left(\operatorname{Tr}[\boldsymbol{\Sigma}]-2\right)=\frac{2 \xi^2}{\Gamma ^2-4 \xi^2+4 \delta \omega^2}=\frac{\xi^2/2}{\xi_c^2-\xi^2}.}
Here and throughout, the derivative defining the QFI is taken with respect to $\delta\omega$ at fixed $\xi,\kappa,\gamma$. The photon number $\bar n$ is used only to parametrize the operating point. To study the near-threshold scaling in the main text, we increase $\xi$ toward $\xi_c$ at fixed $\kappa,\gamma,\delta\omega$, so that,  $\xi=\xi_c \sqrt{\frac{2\nbar}{1+2\nbar}}$.

\subsection{Output QFI}

\begin{figure}[!b]
    \centering
    \includegraphics[width=0.6\linewidth]{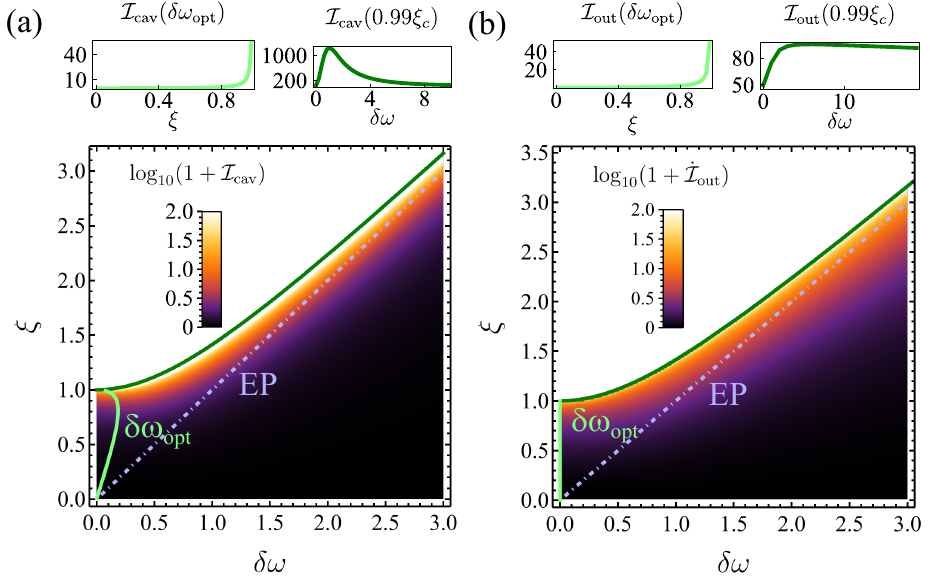}
    \vspace{-0.3cm}
    \caption{\textbf{QFI across OPO parameter space.} (a) Intracavity QFI $\mathcal{I}_{\rm cav}$ and (b) output QFI rate $\dot{\mathcal{I}}_{\rm out}$ as functions of $\delta\omega$ and $\xi$. Top left plots on (a)--(b) show QFI (rate) along the optimal-detuning contours $\delta\omega_{\rm opt}$ (light green), which maximize each quantity at fixed $\xi$. Because $\mathcal{I}_{\rm cav}$ and $\dot{\mathcal{I}}_{\rm out}$ increase continuously as a function of $\xi$ approaching threshold, there is no maximum for a given $\delta\omega$; top right plots on (a)--(b) thus show corresponding quantities at $\xi=0.99\xi_c$. For $\xi>1$, largest QFI (rate) for each $\xi$ is obtained when $\delta\omega_{\rm opt}$ lies close to threshold (not shown). Threshold is shown in solid dark green, EP in dash-dotted purple. The density plot ranges are clipped for visibility. Fixed parameters are $\kappa=\gamma=1$.}
    \label{fig:qfi-supp}
\end{figure}
\subsubsection{Partially Monitored, $\gamma\ne 0$}
We now analyze the near-threshold asymptotic scaling of the output QFI rate for $\bar n\to\infty$, with the steady-state limit $T\to\infty$ taken first as described in \ref{sec:discrete}. Recall the main text definitions
\fla{\xi_c^2=\frac{\Gamma^2}{4}+\delta\omega^2,\qquad
\epsilon&=\xi_c^2-\xi^2,\qquad
\Lambda_\gamma = \delta\omega^2(2\gamma+\kappa)+\frac{\kappa \Gamma^2}{4},\qquad
Q(\omega)=\omega^2\Gamma^2+(\epsilon-\omega^2)^2.}
Using Eq.~\eqref{eq::sm-qfi}, we obtain
\fla{\Ipair(\omega)&=\frac{16 \kappa  \xi ^2 \left(\kappa  \xi_c^2 \omega ^2 \Gamma^2+\Lambda_\gamma  \left(\epsilon -\omega ^2\right)^2\right)}{Q(\omega)^2 \left(2 \gamma  \kappa  \xi ^2+Q(\omega)\right)},\\
\Ipair(0)&=\frac{16 \kappa  \Lambda_\gamma  \xi ^2}{\epsilon ^2 \left(2 \gamma  \kappa  \xi ^2+\epsilon ^2\right)}=\frac{32 \kappa  \Lambda_\gamma  \nbar (2 \nbar+1)^3}{\xi_c^4 \left(4 \gamma  \kappa  (2 \nbar+1) \nbar+\xi_c^2\right)}.}
The quantity $\Ipair(0)\equiv\lim_{\omega\to0^+}\Ipair(\omega)$ is the continuation of the nonzero-frequency pair QFI and is twice the QFI $\mathcal I_0$ of the physical $n=0$ Fourier mode. As $\nbar\to\infty$, we have $\xi^2=\frac{2 \nbar \xi_c^2}{2 \nbar+1} \to \xi_c^2, \; 
\epsilon=\frac{\xi_c^2}{2 \nbar+1} \sim \frac{\xi_c^2}{2\nbar},\;
2\gamma\kappa\xi^2+\epsilon^2 \to 2 \gamma \kappa \xi_c^2$.
Hence $\epsilon\to0^+$ and $\xi\to\xi_c^-$, so $\bar n\to\infty$ approaches threshold from below and
\fla{\Ipair(0)\sim \frac{32 \Lambda_\gamma}{\gamma \xi_c^4}\nbar^2.}
We further have near threshold
$Q(\omega)\to \omega^2 \Gamma^2+\omega^4=\omega^2(\Gamma^2+\omega^2)$,
which is strictly nonzero for $\omega \ne 0$, so the factor that caused divergence at $\omega=0$ no longer vanishes in the denominator. The factors in the numerator of $\Ipair$ approach
$\left(\kappa  \xi_c^2 \omega ^2 \Gamma^2+\Lambda_\gamma  \left(\epsilon -\omega ^2\right)^2\right) \to \omega^2 \left( \kappa \xi_c^2 \Gamma^2 + \Lambda_\gamma \omega^2\right)$.
Explicitly, for a fixed $\omega\ne 0$
\fla{\lim_{\nbar\to\infty}\Ipair(\omega\ne0)&=\frac{16 \kappa \xi_c^2 \left( \kappa \xi_c^2 \Gamma^2 + \Lambda_\gamma \omega^2\right)}{\omega^2(\Gamma^2+\omega^2)^2\left(2\gamma \kappa \xi_c^2+\omega^2(\Gamma^2+\omega^2)\right)},\label{eq::o1}
}
which is finite and independent of $\bar n$ for every fixed $\omega\ne0$. The central frequency region near threshold is controlled by $Q(\omega)\approx \epsilon^2+\Gamma^2\omega^2$. Strong near-threshold dependence is retained for $\omega\lesssim \epsilon/\Gamma \sim \xi_c^2/2\nbar \Gamma=\Delta\omega_{\rm QFI}$. Here $\Delta\omega_{\rm QFI}$ denotes an asymptotic crossover scale rather than an exact linewidth. This does not contradict~\eqref{eq::o1}, which states that for a \textbf{fixed} $\omega$, the critical limit is $\mathcal{O}(1)$.
Since the crossover frequency scales as $\Delta\omega_{\mathrm{QFI}}\sim 1/\bar n$, we set $\omega=z/\bar n$ to resolve the structure of the critical spectral region, so that
\fla{\Ipair\left(\frac{z}{\nbar}\right)\sim\frac{32 \xi_c^2\left(\Lambda_\gamma  \xi_c^2+4 \kappa  z^2 \Gamma^2\right)}{\gamma  \left(\xi_c^4+4 z^2 \Gamma^2\right)^2}\nbar^2=f_\gamma(z)\nbar^2,\label{eq::Ipairzn}
}
and the output QFI rate integrated over all frequencies is
\fla{\dot{\mathcal{I}}_{\rm out}&=\int_0^\infty \frac{d\omega}{2\pi}\Ipair(\omega)\sim\nbar \int_0^\infty \frac{dz}{2\pi}f_\gamma(z)= \nbar\frac{4 \left(\kappa  \Gamma+4 \delta \omega ^2\right)}{\gamma  \left(\Gamma^2+4 \delta \omega ^2\right)}=C_{\nbar} \nbar.\label{eq::qfi-partial}}
The QFI of a near-zero frequency pair scales as $\bar n^2$, but the critically enhanced spectral region narrows as $\frac{1}{\bar n}$. Integration over the stationary spectrum therefore leaves only a linear asymptotic scaling, $\dot{\mathcal I}_{\rm out}\propto\bar n$. Capturing the critical feature requires the finite-time Fourier filter to be narrow enough to sample it, which is equivalent to $\Delta \omega \lesssim  \Delta \omega_{\rm QFI}$, or the measurement record has to satisfy $T\gtrsim \frac{2\pi}{\Delta \omega_{\rm QFI}}$. Replacing the discrete sum by the continuum integral requires the stronger condition $\Delta \omega \ll\Delta \omega_{\rm QFI}\sim\frac{1}{\nbar}$ or $T\gg \frac{2\pi}{\Delta \omega_{\rm QFI}}\sim \nbar$, i.e., each finite-time filter is much narrower than the frequency scale on which the QFI in the critical region changes. The coefficient of the linear scaling satisfies 
\fla{\frac{d}{d(\delta\omega^2)}\left(C_{\nbar}\right)=\frac{16 \Gamma}{\left(\Gamma^2+4 \delta\omega^2\right)^2},}
so that $C_{\nbar}$ increases monotonically with $\delta\omega$ at fixed $\kappa,\gamma$ and has the form
\fla{
C_{\nbar}=\begin{cases}
    \frac{4\kappa}{\gamma\Gamma}& \delta\omega=0,\\
  \frac{4}{\gamma} & \delta\omega / \Gamma\to\infty,
\end{cases}}
Here the large-detuning limit refers to far-detuned operation while maintaining $\bar n\to \infty$, with $\xi$ increased together with $\xi_c$ so that the OPO remains near threshold. Varying $\delta\omega$ while remaining in the critical regime requires $\epsilon\to 0$, and with $\xi=\xi_c\sqrt{\frac{2\nbar}{1+2\nbar}}$, the requirement reduces to $\frac{\delta\omega^2+\Gamma^2/4}{1+2\nbar}\to 0$. For instance, the curve $\delta\omega=\Gamma \nbar^{1/4}$ satisfies this requirement, while also adhering to the far-detuned limit, with $\delta\omega/\Gamma \to \infty$.

We complement Fig. 2  of the main text with a plot of the intracavity QFI and output QFI rate in Fig.~\ref{fig:qfi-supp}.
Along $\xi=0.99\xi_c$ contours, both the output and intracavity QFI initially increase with detuning, but fall off at larger detunings [Figs.~\hyperref[fig:qfi-supp]{S1(a)--S1(b)} top plots], since an operating point increasingly closer to $\xi_c$ is required to remain in the same asymptotic critical regime. One may also track the optimal $\delta\omega_{\rm opt}$ values for every $\xi$ that maximize the QFI (light green lines in Fig.~\ref{fig:qfi-supp}), again highlighting that the optimization of intracavity and output QFI constitute two distinct problems. Nevertheless, we see that both $\mathcal{I}_{\rm cav}$ and $\dot{\mathcal{I}}_{\rm out}$ are enhanced towards threshold, although per photon optimization separates the optimal $\xi$ at each $\delta\omega$ away from threshold as well, which is discussed in the main text.

\subsubsection{Fully Monitored, $\gamma= 0$}

For zero intrinsic loss, $\gamma=0$ and $\Gamma=\kappa$. We then have $\Lambda_0=\kappa \xi_c^2$, and $\kappa \xi_c^2 \omega^2 \Gamma^2+\Lambda_0 (\epsilon-\omega^2)^2 =\kappa \xi_c^2 Q(\omega)$, so the pair QFI reduces to
\fla{
\Ipair(\omega)&=\frac{16\kappa^2\xi_c^4}{Q(\omega)^2},\\
\Ipair(0)&=\frac{16\kappa^2\xi_c^4}{\epsilon^4}\sim\frac{256 \kappa ^2 \nbar^4}{\xi_c^4},}
where again $\Ipair(0)=\lim_{\omega\to0^+}\Ipair(\omega)$. The crossover frequency again scales as $1/\bar n$, so setting $\omega=z/\bar n$ gives
\fla{\Ipair\left(\frac{z}{\nbar}\right)\sim\frac{256 \kappa^2 \xi_c^4}{(4z^2\kappa^2+\xi_c^4)^2}\nbar^4=f_0(z)\nbar^4.}
Thus, unlike the $\gamma>0$ case, the zero-frequency pair QFI scales as $\Ipair(0)\propto\bar n^4$. The integrated spectrum is then
\fla{\dot{\mathcal{I}}_{\rm out}\sim\nbar^3\int_0^\infty \frac{dz}{2\pi}f_0(z)=\frac{64 \kappa }{\kappa ^2+4 \delta \omega ^2}\nbar^3.\label{eq::qfi-fully}}
The coefficient of the QFI rate in the fully monitored OPO is maximized when $\delta\omega=0$, in contrast to a partially monitored OPO, where for fixed $\kappa,\gamma$, the coefficient is maximized in the far-detuned limit. The equation does not imply a super-Heisenberg scaling, since fixed $T$ is incompatible with taking $\nbar \to \infty $ while remaining in the frequency-resolved regime underlying the asymptotic analysis. As discussed above, resolution requires at least $\Delta \omega\lesssim \Delta \omega_{\rm QFI}\sim \frac{\xi_c^2}{2\nbar \Gamma}$ (with $\gamma=0$, $\frac{\xi_{c}^2}{2\nbar \kappa}$), or, equivalently, $T\gtrsim \frac{2\pi}{\Delta \omega_{\rm QFI}} \sim \nbar$, which itself grows as threshold is approached.
Then, the output QFI scales as
\fla{\mathcal{I}_{\rm out}&\sim\nbar T \frac{4}{\gamma}\frac{\kappa \Gamma/4+\delta\omega^2}{\Gamma^2/4+\delta\omega^2} \le \frac{4}{\gamma}\nbar T=\frac{4}{\gamma \kappa}N_{\rm out},\qquad \gamma> 0,\\
\mathcal{I}_{\rm out}&\sim\nbar^3 T\frac{16\kappa}{\kappa^2/4+\delta\omega^2}= \frac{16 \kappa}{\xi_{c}^2}\nbar^3 T,\qquad \gamma=0,}
Rearranging, we get 
\fla{\mathcal{I}_{\rm out}&=\mathcal{O}(\nbar T)=\mathcal{O}(N_{\rm out}),\qquad \gamma> 0,\\
\mathcal{I}_{\rm out}&=\mathcal{O}(\nbar^3 T)\lesssim\mathcal{O}(\nbar^2T^2)=\mathcal{O}(N_{\rm out}^2),\qquad \gamma=0,}
where $N_{\rm out}= \kappa \nbar T$ is the emitted photon number in the monitored channel for a steady-state record, and we used $T\gtrsim \frac{2\pi}{\Delta\omega_{\rm QFI}}\sim \nbar$. Thus, even the minimal requirement for resolving the critical feature prevents scaling faster than $N_{\rm out}^2$. Moreover, replacing the discrete frequency sum by the continuum integral used in Eq.~\eqref{eq::qfi-fully} requires the stronger condition $T/\nbar\to \infty$ so that $\mathcal{I}_{\rm out}/N_{\rm out}^2\sim\nbar/T\to 0$, and the quadratic scaling is not achieved within our treatment.

\subsubsection{Fundamental Metrological Bound on the Steady-State Output QFI Rate}
Following the analysis in Refs.~\cite{alushi2024optimality,gualrecki2025time} we apply the adaptive metrological bound of Ref.~\cite{demkowicz2017adaptive} to our partially monitored OPO. For a quantum state $\rho$ described by the Lindblad equation
\fla{\frac{d\rho}{dt}=-i\delta\omega[H,\rho]+\sum_{j=1}^{J}\left(L_j \rho L_j^\dag-\frac{1}{2}\{L_j^\dag L_j,\rho\}\right),}
where $\delta\omega$ is estimated and $H=\partial_{\delta\omega}\mathcal{H}$ is the generator part of the full Hamiltonian,
one can define~\cite{demkowicz2017adaptive}
\fla{\beta^{(1)}&=H+h_{00}^{(1)}\mathbb{1}+\mathbf{h}^{\dag (\frac{1}{2})}\mathbf{L}+\mathbf{L}^\dag \mathbf{h}^{(\frac{1}{2})}+\mathbf{L}^\dag \mathfrak{h}^{(0)}\mathbf{L},\\
\alpha^{(1)}&=(\mathbf{h}^{(\frac{1}{2})}\mathbb{1}+\mathfrak{h}^{(0)}\mathbf{L})^\dag(\mathbf{h}^{(\frac{1}{2})}\mathbb{1}+\mathfrak{h}^{(0)}\mathbf{L}),}
where $\mathbf{L}$ is a vector of Lindblad operators, $h_{00}^{(1)}$ is a real scalar, $\mathbf{h}^{(\frac{1}{2})}$ is a vector of length $J$ and $\mathfrak{h}^{(0)}$ is a $J \times J$ Hermitian matrix
parametrizing the freedom in the Kraus representation. If $\beta^{(1)}$ can be chosen to vanish, the QFI of any adaptive protocol with arbitrary parameter-independent controls and ancillas obeys
\fla{\mathcal{I} \le 4 \int_0^T \underbrace{\min}_{h}\mean{\alpha^{(1)}}_t dt \qquad \text{subject to } \beta ^{(1)}=0.}
Otherwise, a scaling quadratic in $T$ is not ruled out. We apply this framework to the partially monitored OPO by taking the infinitesimal sensing channel to contain the detuning evolution and unmonitored intrinsic loss 
\fla{\mathcal{E}^{\delta\omega}_{\Delta t}(\rho)=\rho+(-i\delta\omega[H,\rho]+\gamma {\mathcal{D}}[a]\rho)\Delta t+\mathcal{O}(\Delta t^2),}
where ${\mathcal{D}}[a]\rho=a\rho a^\dag - \frac{1}{2}\{a^\dag a,\rho\}$ is the intrinsic loss part of the master equation. The ancilla states are the discretized temporal input-output modes
\fla{b_{j,\rm in/out}=\frac{1}{\sqrt{\Delta t}}\int_{t_j}^{t_j+\Delta t} A_{\rm in/out}(t)dt,}
with inputs initially in vacuum. After interacting with the cavity they constitute the corresponding monitored output modes. The control map $\mathcal{C}_j$ is applied after each infinitesimal $\Delta t$ and is composed of parameter-independent squeezing and a beamsplitter interaction between the input/output and cavity $\mathcal{C}_j(\rho)=U_{\kappa,j}U_{\xi,j}(\rho)U^\dag_{\xi,j}U_{\kappa,j}^\dag$,
where 
\fla{U_{\kappa,j}=\exp\left(\sqrt{\kappa \Delta t} (a^\dag b_j-a b_j^\dag) \right),\qquad
U_{\xi,j}=\exp\left(\frac{\xi\Delta t}{2}({a^\dag}^2-a^2)\right).}
Their interleaving with $\mathcal{E}_{\Delta t}^{\delta\omega}$ reproduces the full OPO dynamics as $ \Delta t\to0$. Thus, the protocol has the form assumed by the theorem. We now apply the theorem to the QFI generated during a steady-state measurement interval $T$, excluding information accumulated during preparation. Since the accessible state includes the cavity and monitored $\kappa$ output, the only unmonitored Lindblad operator is $L=\sqrt\gamma\,a$, while $H=a^\dag a$. Choosing $h_{00}^{(1)}=\mathbf h^{(1/2)}=0$ gives
\fla{\beta^{(1)}=a^\dag a+\mathfrak{h}^{(0)}\gamma a^\dag a \quad \longrightarrow\quad \mathfrak{h}^{(0)}=-\frac{1}{\gamma} \quad \longrightarrow\quad \alpha^{(1)}=\frac{1}{\gamma^2}L^\dag L=\frac{a^\dag a}{\gamma} .}
We apply the theorem to the QFI generated during a steady-state interval $T$, excluding information accumulated during preparation. The monitored output stationary information rate cannot exceed the joint cavity-monitored-output information growth in steady-state $\mathcal{I}_{\rm cav+out}$
\fla{\mathcal{I}_{\rm out}\le\mathcal{I}_{\rm cav+out}\le \frac{4}{\gamma}\int_0^T \mean{a^\dag a}_t dt=\frac{4 \nbar T}{\gamma},} 
where $t=0$ is assumed to indicate the start of measurement, after steady-state is established.  When $\gamma=0$, there are no unmonitored Lindblad operators and $\beta^{(1)}=a^\dag a+h_{00}^{(1)}\mathbb{1}$ cannot vanish. Hence the theorem provides no linear-in-$T$ bound on the total QFI. In the previous section we nevertheless argued that the scaling quadratic in $N_{\rm out}$ is not achieved within our treatment. 

The preparation contribution omitted above is also bounded by the same theorem. For $\xi\ge\delta\omega$, define the characteristic time to establish a steady-state near threshold based on the slow eigenvalue $\lambda_+$ of $\boldsymbol{\mathcal{E}}$, $\tau_c\sim|\operatorname{Re}\lambda_+|^{-1}=\left(\Gamma/2-\sqrt{\xi^2-\delta\omega^2}\right)^{-1}\sim\frac{2\Gamma \nbar}{\xi_c^2}\sim \nbar$. Since $\mean{a^\dag a}_t=\mathcal{O}(\nbar)$, then output QFI during transient dynamics $\mathcal{I}_{\rm prep}$ is upper bounded by $\mathcal{I}_{\rm prep}\le \frac{4}{\gamma}\int_{\rm prep} dt \; \mean{a^\dag a}_t=\mathcal{O}(\nbar \tau_c/\gamma)$. Hence, under the continuum condition $T/\bar n\to\infty$, $\mathcal I_{\rm prep}/\mathcal I_{\rm out} = \mathcal{O}\left(\tau_c/T\right)\to0$. A one-time preparation therefore does not modify the leading steady-state asymptotics. 
For $\xi<\delta\omega$, the relaxation time is $\mathcal{O}(\Gamma^{-1})$ and therefore gives an even smaller preparation contribution.

We showed previously that when $\delta\omega/\Gamma\gg1$, the output QFI approaches $4\nbar T/\gamma$, so the output alone exhausts the joint upper bound at leading order for a measurement record of duration $T$ in steady-state. To see all the parametric regimes saturating the bound, consider the quantity $\gamma\dot{\mathcal{I}}_{\rm out}/4\nbar=1-\frac{\gamma/\Gamma}{1+4(\delta\omega/\Gamma)^2}$ for finite $\gamma$, which approaches $1$ either when $\delta\omega/\Gamma\to \infty$ or $\gamma/\Gamma\to0 \longleftrightarrow \kappa/\gamma \to \infty$. The two limits may be taken simultaneously. At the same time, the steady-state intracavity QFI for fixed $\delta\omega\ne 0$ (the intracavity QFI is linear in $\nbar$ when $\delta\omega=0$)
\fla{\mathcal{I}_{\rm cav}\sim \frac{128 \delta\omega^2}{(\Gamma^2+4\delta\omega^2)^2}\nbar^2\sim\frac{4\delta\omega^2}{\Gamma \xi_c^2} \nbar \tau_c.}
At fixed $\nbar$, the leading term is maximized at $\delta\omega=\Gamma/2$, equivalent to the optimum found in Ref.~\cite{alushi2024optimality}. If only the intracavity state is accessible, so both decay channels are unmonitored and $L_1=\sqrt{\gamma} a$ and $L_2=\sqrt{\kappa}a$, we would obtain for the fundamental bound on the intracavity QFI, as in~\cite{alushi2024optimality}
\fla{\mathcal{I}_{\rm cav}\le \frac{4}{\Gamma}\int_0^T \mean{a^\dag a}_t dt\lesssim\frac{4\nbar T}{\Gamma},}
where we include the transient dynamics in the measurement record, which allows only for a loose bound. A single steady-state intracavity snapshot requires a preparation time of order $\tau_c$, and continuing the evolution after the steady state is established does not increase its QFI. Thus, a loose upper envelope for the snapshot QFI is $\mathcal I_{\rm cav}\lesssim4\bar n\tau_c/\Gamma$. At the single-shot optimum $\delta\omega=\Gamma/2$, $\mathcal I_{\rm cav}\sim2\bar n\tau_c/\Gamma$. Therefore, repeating the preparation and interrogation over $\mathcal{O}(T/\tau_c)$ cycles would recover the same $\mathcal{O}(\bar nT)$ scaling.
At large detuning, $\mathcal{I}_{\rm out}$ approaches the joint bound for a single continuous measurement with $T\gg\tau_c$. By contrast, the QFI of a single steady-state intracavity snapshot remains $\mathcal{O}(\bar n\tau_c)$, so $\mathcal{I}_{\rm cav}/\mathcal{I}_{\rm out}\sim \gamma\tau_c/\Gamma T\to0$. This is not to suggest that the cavity-output QFI is obtained as a sum of individual components, since information is also encoded in their correlations. However, access to the intracavity snapshot provides no additional leading-order information in a regime where the output saturates the joint bound.

\subsection{Quantum Fisher Information at the Output of a Passive Sensor}
We compare the OPO with a passive cavity ($\xi=0$) driven by a monochromatic coherent field $\langle A_{\rm in}(t)\rangle=\alpha_d e^{-i\omega_dt}$. The passive-cavity equation of motion in the frequency domain is
\fla{\begin{pmatrix}
    x(\omega) \\ p(\omega)
\end{pmatrix}=\begin{pmatrix}
    \frac{\Gamma}{2} & \omega-\delta \omega \\
    \delta \omega - \omega &   \frac{\Gamma}{2}
\end{pmatrix}^{-1}  \left[ \sqrt{\kappa} \begin{pmatrix}
    X_{\rm in}(\omega) \\ P_{\rm in}(\omega)
\end{pmatrix}+\sqrt{\gamma} \begin{pmatrix}
    X_{\rm in}^{(B)}(\omega) \\ P_{\rm in}^{(B)}(\omega)
\end{pmatrix}\right] .}
With quadrature Green function $\mathbf G_R(\omega)$, the input-output relation becomes $\textbf{R}_{\rm out}=(\mathbb{1}-\kappa \textbf{G}_R) \textbf{R}_{\rm in}-\sqrt{\kappa \gamma} \mathbf{G}_R\textbf{R}_{\rm in}^{(B)}$. The output covariance is
\fla{\Sout=(\mathbb{1}-\kappa \mathbf{G}_R)\boldsymbol{\Sigma}_{\rm in,R}(\mathbb{1}-\kappa \mathbf{G}_R)^T + \kappa \gamma \mathbf{G}_R \boldsymbol{\Sigma}_{\rm in,R}^{(B)}\mathbf{G}_R^T=\mathbb{1}_2,}
with $\boldsymbol{\Sigma}_{\rm in,R}=\boldsymbol{\Sigma}_{\rm in,R}^{(B)}=\mathbb{1}_2$ for a coherent input and vacuum intrinsic-loss bath.  
The mean intracavity photon number is 
$\nbar=\frac{\kappa|\alpha_d|^2}{\Gamma^2/4+(\delta \omega-\omega_d)^2}$.
The finite-duration discrete frequency mode is $A_{\rm in/out,n}=\frac{1}{\sqrt{T}}\int_{t_{j}}^{t_j+T} dt \; e^{i \omega_n t} A_{\rm in/out}(t)$ as before, so that $\mean{A_{\rm in}(t)}=\alpha_d e^{-i \omega_d t}$ and $\mean{A_{\rm in,n}}=\sqrt{T} \alpha_d \delta_{n,n_d}$ where we take $\omega_d=\omega_{n_d}$. In the long-time limit, the covariance assigned to each discrete frequency mode approaches $\boldsymbol{\Sigma}(\omega_n)=\mathbb{1}_2$, and only the displacement contribution to the single-mode Gaussian QFI survives
\fla{\mathcal{I}_{\rm passive}(\omega_d)=2{\mathbf{d}'}_{\rm out,n_d}^T\boldsymbol{\Sigma}_{\rm out,n_d}^{-1} {\mathbf{d}'}_{\rm out,n_d}=\frac{4 \kappa }{\Gamma^2/4+ (\delta \omega -\omega_d )^2}\nbar T,}
where the derivative is taken with respect to $\delta\omega$ at fixed $\alpha_d,\omega_d,\kappa,\gamma$, and $\bar n$ parametrizes the operating point. The passive cavity with a coherent state can then saturate the QFI bound in steady-state with $\omega_d=\delta\omega$ and $\kappa=\gamma$, for which $\mathcal{I}_{\rm passive}=\frac{4\nbar T}{\gamma}$.

\section{Output Squeezing and Homodyne Fisher Information}

\subsection{Output Squeezing Spectrum}

\begin{figure}[!b]
    \centering
    \includegraphics[width=0.6\linewidth]{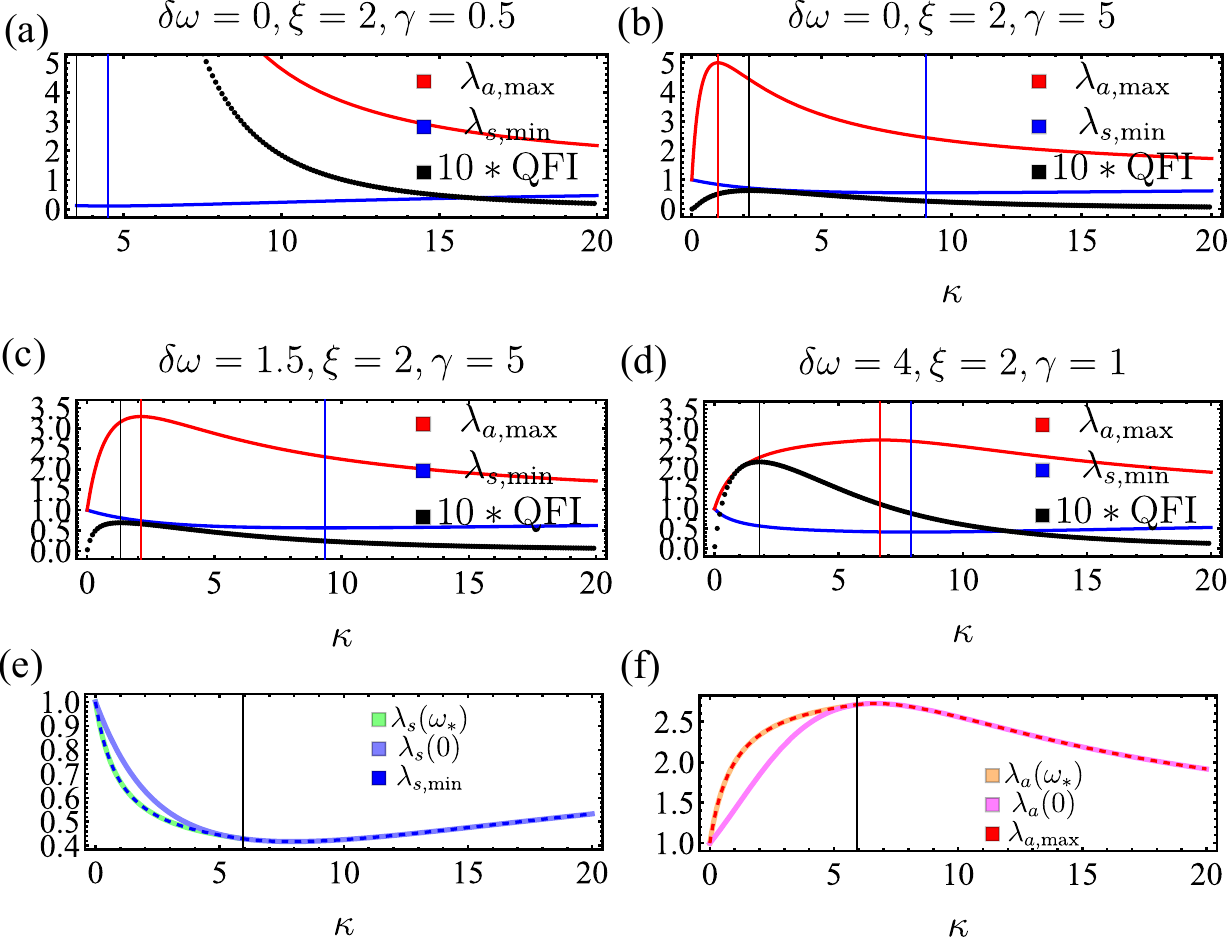}
    \vspace{-0.3cm}
    \caption{\textbf{Optimal outcoupling for maximum squeezing/antisqueezing compared with that maximizing the QFI.} (a) $\delta\omega=0$; the threshold is reached at a finite positive $\kappa$, where the QFI and antisqueezing diverge, whereas squeezing exhibits a finite optimum. (b) $\delta\omega=0$ with sufficiently large intrinsic loss, where the threshold cannot be reached for any physical $\kappa$, and all three quantities develop an optimal outcoupling condition. (c) Same as (b) but for $\delta\omega\ne 0$ in the single-peaked squeezing spectrum regime $\delta \omega^2 < \xi^2+\Gamma^2/4$. (d) $\delta\omega\ne 0$ in the double-peaked spectrum regime $\delta \omega^2 > \xi^2+\Gamma^2/4$ that becomes single-peaked as $\kappa$ increases. Maximal squeezing/antisqueezing initially occur at $\omega_*$ and do not exhibit an optimal outcoupling; as $\kappa$ increases, the most squeezed/antisqueezed frequency moves to $\omega=0$, after which an optimum develops. (e) Squeezed eigenvalues corresponding to (d), showing the frequency of maximal squeezing moving from the $\omega_*$ sideband to $\omega=0$. (f) Antisqueezed eigenvalue corresponding to (d), showing analogous crossover. Extrema on all plots are shown with vertical lines.}
    \label{fig2:supp}
\end{figure}
In the long-time limit, the output covariance is block diagonal in correlated $\pm\omega$ sectors. The doubly degenerate eigenvalues $\lambda_s(\omega)<\lambda_a(\omega)$ of each $4\times4$ covariance block give the minimum and maximum quadrature variances, respectively, and therefore define the squeezing and antisqueezing spectra. Both spectra have stationary points at $\omega=0$ and, when real, at $\omega=\omega_*=\pm \sqrt{\delta\omega^2-\xi^2-\Gamma^2/4}$. Thus the spectrum is single-peaked for $\delta\omega^2<\xi^2+\Gamma^2/4$, while for $\delta\omega^2>\xi^2+\Gamma^2/4$, the optimal squeezing and antisqueezing occur at the finite-frequency sidebands $\pm\omega_*$. The eigenvalues at the peaks are
\fla{\lambda_{s,\min}&=\begin{cases}
1-\frac{8 \kappa  \xi }{\sqrt{16 \xi ^2 \Gamma^2+16 \left(\xi_c^2-\xi ^2\right)^2}+4 \xi  \Gamma} & \delta\omega^2<\xi^2+\Gamma^2/4\\
    \frac{\gamma}{\Gamma}+\frac{\kappa}{\Gamma} \frac{\delta \omega - \xi}{\delta \omega + \xi} & \delta\omega^2>\xi^2+\Gamma^2/4
\end{cases},\label{eq::sq}\\
\lambda_{a,\max}&=\begin{cases}
    1+\frac{8 \kappa  \xi }{\sqrt{16 \xi ^2 \Gamma^2+16 \left(\xi_c^2-\xi ^2\right)^2}-4 \xi \Gamma} & \delta\omega^2<\xi^2+\Gamma^2/4\\
    \frac{\gamma}{\Gamma}+\frac{\kappa}{\Gamma} \frac{\delta \omega +\xi}{\delta \omega - \xi}& \delta\omega^2>\xi^2+\Gamma^2/4
\end{cases}\label{eq::as}.}
In the split regime, $\omega_*$ is a local minimum of $\lambda_s(\omega)$ and a local maximum of $\lambda_a(\omega)$, while $\omega=0$ has the opposite character. At fixed $\kappa,\gamma,\delta\omega$, direct differentiation of Eqs.~\eqref{eq::sq}-\eqref{eq::as} shows that $\partial_\xi\lambda_{s,\min}<0$ and $\partial_\xi\lambda_{a,\max}>0$ throughout both spectral regimes. Hence threshold is globally optimal with respect to $\xi$ for both squeezing and antisqueezing, with $\lambda_{s,\min} \to \gamma/\Gamma$ and $\lambda_{a,\max}\to \infty$. 

To compare with the QFI spectrum, we examine the near-threshold scaling of the squeezing and antisqueezing spectra. Expressing, $\lambda_a(\omega)=1+\frac{8 \kappa  \xi }{R(\omega )-4 \xi  \Gamma}$ and $\lambda_s(\omega)=1-\frac{8 \kappa  \xi }{4 \xi  \Gamma+R(\omega )}$ with $R(\omega)=4\sqrt{\xi ^2 (\gamma +\kappa )^2+Q(\omega )}$, we examine the critical spectral region with $\omega\sim \frac{z}{\nbar}$, so that $R\left(\frac{z}{\nbar}\right)\sim 4\Gamma\xi+\frac{\xi_c^4+4\Gamma^2z^2}{2 \nbar^2\Gamma \xi_c}$ near threshold. The asymptotic expressions for $\lambda_{s,a}(\omega)$ in the critical region near threshold are
\fla{\lambda_s\left(\frac{z}{\nbar}\right)&\sim\frac{\gamma}{\Gamma}+\frac{\kappa}{\nbar^2}\frac{\xi_c^4+4\Gamma^2z^2}{16 \Gamma^3 \xi_c^2},\\
\lambda_a\left(\frac{z}{\nbar}\right)&\sim \frac{16\kappa \Gamma}{\xi_c^2} \frac{\nbar^2}{1+4\Gamma^2 z^2/\xi_c^4}.} 
Thus the critical spectral region narrows as $1/\bar n$. Within this region the antisqueezing peak grows as $\bar n^2$, while the squeezed variance approaches its minimum value $\gamma/\Gamma$ with an $\mathcal{O}(\bar n^{-2})$ correction. By contrast, at any fixed $\omega\ne0$, both $\lambda_s(\omega)$ and $\lambda_a(\omega)$ remain $\mathcal{O}(1)$ as $\bar n\to\infty$, so the divergent antisqueezing is confined to the same narrowing critical spectral region, while $\lambda_s(0)\sim\frac{\gamma}{\Gamma}+\frac{\kappa\xi_c^2}{16 \Gamma^3 \nbar^2}$ and $\lambda_a(0)\sim \frac{16 \kappa \Gamma}{\xi_c^2}\nbar^2$.

When the most squeezed frequency is in the resolved sidebands, there is no extremum with respect to $\kappa$ for that nonzero frequency mode, since the first derivative with respect to $\kappa$ of $\lambda_s(\omega_*)$ and $\lambda_a(\omega_*)$ is $\mp\frac{2 \gamma  \xi }{\Gamma^2 (\delta \omega \pm \xi )}$, respectively, so that $\lambda_s(\omega_*)$ decreases monotonically and $\lambda_a(\omega_*)$ increases monotonically with $\kappa$. An optimum develops only once increasing $\kappa$ shifts the frequency of maximal squeezing/antisqueezing back to $\omega=0$ [Figs.~\hyperref[fig2:supp]{S2(d)--S2(f)}]. At $\delta\omega=0$, the optimal outcoupling for maximal squeezing is $\kappa=\gamma+2\xi$, and for maximal antisqueezing is $\kappa=\gamma-2\xi$. The latter is physical only for $\gamma>2\xi$, i.e., when the intrinsic loss alone is sufficient to keep the system below threshold at $\kappa=0$ [Figs.~\hyperref[fig2:supp]{S2(a)--S2(b)}]. More generally, for $\delta\omega<\xi$, this condition becomes $\gamma>2\sqrt{\xi^2-\delta \omega^2}$, while the corresponding optimal $\kappa$ for $\delta\omega \ne 0$ is then found numerically [Fig.~\hyperref[fig2:supp]{S2(c)}].

\begin{figure}[t]
    \centering
    \includegraphics[width=0.6\linewidth]{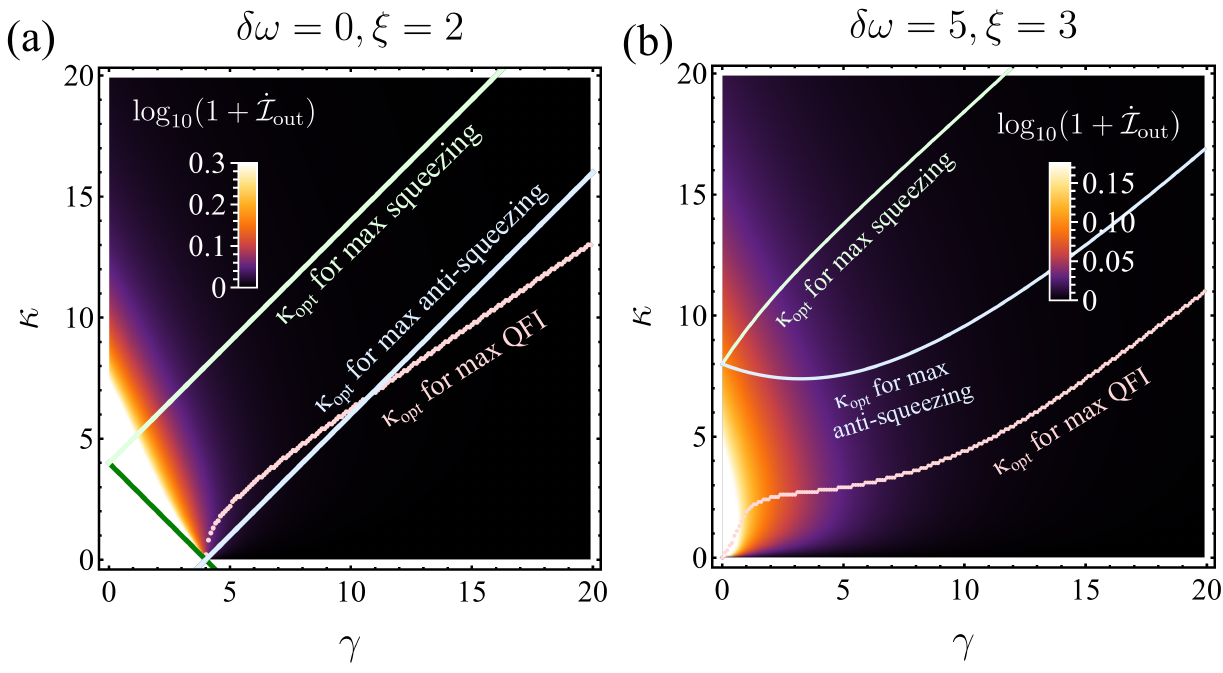}
    \vspace{-0.3cm}
    \caption{\textbf{Output QFI rate and optimal outcouplings across $\kappa-\gamma$ parameter space}. Output QFI rate $\dot{\mathcal{I}}_{\rm out}$, together with the outcoupling $\kappa_{\rm opt}$ maximizing the QFI, squeezing, and antisqueezing. (a) $\xi=2$ and $\delta\omega=0$, for which the squeezing spectrum is single-peaked. (b) $\xi=3$ and $\delta\omega=5$, for which the spectrum crosses from the double-peaked regime, $\delta\omega^2>\xi^2+\Gamma^2/4$, to the single-peaked regime as $\Gamma$ increases.}
    \label{fig3:supp}
\end{figure}

Optimizing squeezing is therefore not equivalent to optimizing metrological information. As shown in Fig.~\ref{fig3:supp} the outcoupling values minimizing $\lambda_{s,\min}$, maximizing $\lambda_{a,\max}$, and maximizing the output QFI rate generally differ. This occurs both when the spectrum remains single-peaked [Fig.~\hyperref[fig3:supp]{S3(a)}] and when varying $\kappa$ drives the spectrum between the double- and single-peaked regimes [Fig.~\hyperref[fig3:supp]{S3(b)}]. The reason becomes clear in the supermode phase-space representation developed next. A detuning perturbation changes both the principal variances and the orientation of the covariance ellipse, whereas squeezing and antisqueezing characterize only the width of its principal axes.

\subsection{Wigner Function}
The duals of the covariance eigenvectors corresponding to the doubly degenerate eigenvalues $\lambda_a(\omega)$ and $\lambda_s(\omega)$ define, up to normalization, the quadratures associated with the principal phase-space directions,
\fla{
\lambda_a(\omega)&: \qquad e^{i\thet} \Atout+e^{-i\thet} \Aout^\dag, \qquad e^{i\thet} \Aout+e^{-i\thet} \Atout^\dag \\
\lambda_s(\omega)&: \qquad  -i e^{i\thet} \Aout+i e^{-i\thet} \Atout^\dag, \qquad -i e^{i\thet} \Atout + i e^{-i\thet} \Aout^\dag}
where $\theta = \operatorname{Arg}\left(\frac{\Gamma^2}{4}+\xi ^2+\omega ^2-\delta \omega ^2+i\delta \omega  \Gamma\right)$.
To visualize the Wigner function in two-dimensional phase space, we introduce the symmetric and antisymmetric sideband supermodes $B_{\pm}=\frac{\Aout\pm \Atout}{\sqrt{2}}$, and $B^\dag_\pm=\frac{\Aout^\dag\pm \Atout^\dag}{\sqrt{2}}$, so that $B_{\pm}=\frac{X_{\pm}+ i P_\pm}{\sqrt{2}}$, $X_\pm=\frac{\Xout\pm \Xtout}{\sqrt{2}}$ and $P_\pm=\frac{\Pout\pm \Ptout}{\sqrt{2}}$. A general quadrature of either supermode is denoted by 
\fla{Q_{\pm,\beta}=X_\pm \cos \beta + P_\pm \sin \beta = \frac{e^{-i\beta} B_\pm + e^{i\beta} B_\pm^\dag}{\sqrt{2}}.}
Symmetric and antisymmetric combinations of quadratures corresponding to each $\lambda_s(\omega)$ and $\lambda_a(\omega)$ give the principal quadratures of the two supermodes, summarized in Table~\ref{tab:principal_quadratures}.
\begin{table}[t]
\centering
\begin{tabular}{c|cc}
 & Antisqueezed quadrature & Squeezed quadrature \\
\hline
$+$ &
$X_+\cos\thet-P_+\sin\thet$ &
$X_+\sin\thet+P_+\cos\thet$ \\[2mm]
$-$ &
$X_-\sin\thet+P_-\cos\thet$ &
$X_-\cos\thet-P_-\sin\thet$
\end{tabular}
\caption{Principal quadratures of the symmetric and antisymmetric output supermodes.}
\label{tab:principal_quadratures}
\end{table}
Thus, the two supermodes have the same principal variances, but their squeezed and antisqueezed axes are interchanged. As established previously, the supermodes are uncorrelated, so the joint covariance matrix is block diagonal and the zero-mean Gaussian state factorizes between them. Their Wigner functions are therefore
\fla{W(X_{\pm},P_{\pm})=\frac{\exp \left( - \begin{pmatrix}
    X_{\pm} & P_{\pm}
\end{pmatrix} \boldsymbol{\Sigma}_{\pm}^{-1}\begin{pmatrix}
    X_{\pm} \\ P_{\pm}
\end{pmatrix}\right)}{\pi \sqrt{\lambda_a \lambda_s}},}
where $\boldsymbol{\Sigma}_+^{-1}=\mathbf{M}^T\begin{pmatrix}
    \lambda_a^{-1} & 0 \\ 0 & \lambda_s^{-1}
\end{pmatrix} \mathbf{M}$ and $\boldsymbol{\Sigma}_-^{-1}=\mathbf{M}^T\begin{pmatrix}
    \lambda_s^{-1} & 0 \\ 0 & \lambda_a^{-1}
\end{pmatrix} \mathbf{M}$ with $\mathbf{M}=\begin{pmatrix}
    \cos \thet & -\sin \thet \\\sin\thet & \cos \thet
\end{pmatrix}$. 
In the main text, we plot only the $+$ supermode plane. At $\omega=0$, there is only one physical mode, as discussed in \ref{SMR}; the corresponding single-mode Wigner function is obtained from the same expression by identifying $X_+\to X$ and $P_+\to P$.

\subsection{Homodyne Detection}

\subsubsection{Photocurrent Measurement}
For homodyne detection with LO phase $\phi/2$, the Fourier component of the photocurrent is the Fourier transform of the time-domain homodyne quadrature~\cite{lvovsky2015squeezed},
\fla{I_\phi(\omega) &\sim \left( e^{-i\ph} A_{\rm out}(\omega)+e^{i\ph} A_{\rm out}^\dag(\omega)\right),\\
u_1&=\operatorname{Re}(I_\phi)=X_+\cos \ph + P_+ \sin \ph = Q_{+,\ph}, \\
u_2&=\operatorname{Im}(I_{\phi})=-X_{-}\sin \ph + P_{-} \cos \ph = Q_{-,\ph+\frac{\pi}{2}}.}
Thus a single LO phase simultaneously measures one quadrature of each supermode. This construction applies for $\omega>0$. The orthogonal quadrature is $I_{\phi+\pi}\left(\omega\right)\sim -i \left( e^{-i \ph} A_{\rm out}(\omega) - e^{i \ph} A_{\rm out}^\dag (\omega)\right)$.
To transform the covariance, we complete $u_1,u_2$ with the orthogonal quadratures into the basis $\mathbf{u}_\phi=\mathbf{T}\mathbf{R}_{\rm out}$, where
\fla{\mathbf{u}_\phi &= \left( \begin{array}{c}
   Q_{+,\ph} \\ Q_{-,\ph+\frac{\pi}{2}} \\ Q_{+,\ph+\frac{\pi}{2}} \\ -Q_{-,\ph}
\end{array}\right)=\frac{1}{\sqrt{2}}\left(
\begin{array}{cccc}
 \cos \left(\frac{\phi }{2}\right) & \cos \left(\frac{\phi }{2}\right) & \sin \left(\frac{\phi }{2}\right) & \sin \left(\frac{\phi }{2}\right) \\
 -\sin \left(\frac{\phi }{2}\right) & \sin \left(\frac{\phi }{2}\right) & \cos \left(\frac{\phi }{2}\right) & -\cos \left(\frac{\phi }{2}\right) \\
 -\sin \left(\frac{\phi }{2}\right) & -\sin \left(\frac{\phi }{2}\right) & \cos \left(\frac{\phi }{2}\right) & \cos \left(\frac{\phi }{2}\right) \\
 -\cos \left(\frac{\phi }{2}\right) & \cos \left(\frac{\phi }{2}\right) & -\sin \left(\frac{\phi }{2}\right) & \sin \left(\frac{\phi }{2}\right) \\
\end{array}
\right)\begin{pmatrix}
    \Xout \\ \Xtout \\ \Pout \\ \Ptout
\end{pmatrix}.}
The first two components have a zero-mean Gaussian joint distribution with vacuum-normalized covariance $\tilde\Sigma_{2\times2}=\mathrm{diag}(V_1,V_2)$, where $\tilde\Sigma=\mathbf T\Sout\mathbf T^T$ and $\operatorname{Var}(u_i)=V_i/2$. Direct evaluation gives $V_1=V_2\equiv V$.
Since $u_1$ and $u_2$ are independent normal variables, the classical Fisher information (CFI) of a multivariate Gaussian~\cite{malago2015information} decomposes as
\fla{\mathcal{F}_{\rm pair}(\omega,\phi)=\frac{1}{2}\operatorname{Tr}\left[\left(\tilde{\Sigma}_{2\times2}^{-1} \partial_{\delta\omega} \tilde{\Sigma}_{2\times2}\right)^2\right]=\frac{1}{2}\left(\frac{\partial_{\delta\omega} V_1}{V_1} \right)^2+\frac{1}{2}\left(\frac{\partial_{\delta\omega} V_2}{V_2} \right)^2.}
From the principal-axis representation obtained above, the common variance is
\fla{V=\frac{\lambda_s+\lambda_a}{2}+\frac{\lambda_a-\lambda_s}{2} \cos\left( \phi+\theta\right)=\lambda_s \sin^2\left(\frac{\theta+\phi}{2} \right)+\lambda_a \cos^2\left(\frac{\theta+\phi}{2} \right),}
where $\ph$ is the angle of the LO and $\theta$ is the angle associated with the Wigner function. Note that the CFI is degenerate for angles $\theta+\phi$ and $-\theta-\phi$ when $\delta\omega=0$. 
At $\delta\omega=0$, $\lambda_{s,a}'=0$, so the first-order change in $V$ arises entirely from rotation of the covariance ellipse. Up to phase-independent factors,
\fla{\mathcal{F}_{\rm pair}(\omega,\phi)\sim \left( \frac{-\frac{\lambda_a-\lambda_s}{2}\sin(\phi)}{\lambda_a \cos^2\left(\ph\right)+\lambda_s \sin^2\left(\ph\right)}\right)^2,}
which is maximized at $\phi_{\rm opt}=\pm2\arctan\sqrt{\lambda_a/\lambda_s}$. For $\lambda_a\gg\lambda_s$, the optimal quadrature approaches the squeezed axis. This reflects the fact that the CFI depends on the fractional variance change, so a narrow marginal distribution enhances distinguishability.
\begin{figure}[!t]
    \centering
    \includegraphics[width=0.5\linewidth]{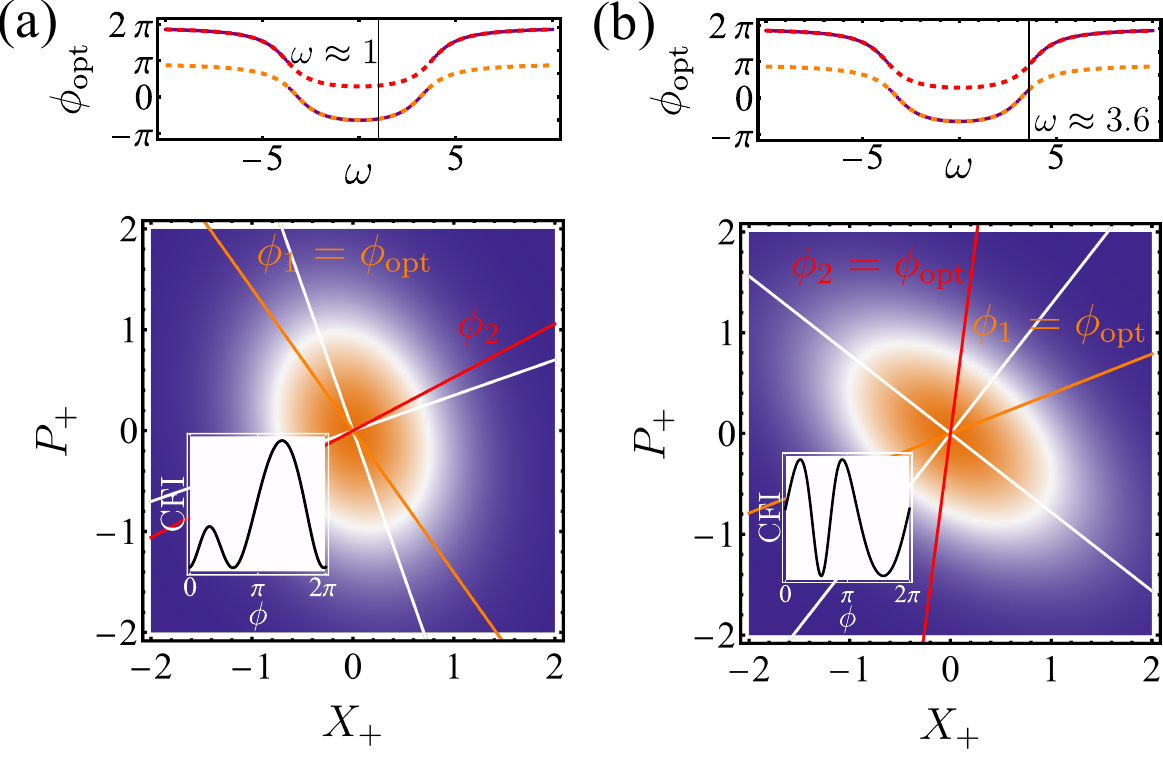}
    \vspace{-0.3cm}
    \caption{\textbf{Homodyne detection for a nonzero-detuning OPO}. The two homodyne-angle solutions corresponding to the largest positive and negative fractional changes in variance are shown in orange and red, respectively, in the top plots of (a)--(b). (a) Wigner function $W(X_{+},P_+)$ at $\omega=1$ where $\phi_1=\phi_{\rm opt}$ (largest negative fractional variance change direction). (b) $W(X_{+},P_+)$ plot at $\omega\approx 3.6$, which is the frequency where the globally optimal quadrature angle switches between the two solutions. Parameters: $\delta\omega=4,\; \xi=2,\; \gamma=\kappa=1$.}
    \label{fig4:supp}
\end{figure}
When $\delta\omega\ne0$, the derivative of the marginal variances is
\fla{\partial_{\delta\omega}V=\lambda_a'\cos^2 \left( \frac{\theta+\phi}{2}\right)+\lambda_s'\sin^2 \left( \frac{\theta+\phi}{2}\right)+(\lambda_s-\lambda_a)\sin(\theta+\phi)\frac{\theta'}{2}.}
We identify the first two terms as the stretching contribution $\Lambda$ and the last as the rotation contribution $\mathcal{R}$. Under $\theta+\phi\to-\theta-\phi$, the stretching contribution $\Lambda$ is even whereas the rotation contribution $\mathcal{R}$ is odd. Hence the two mirror-related fractional variance changes are proportional to $\Lambda+\mathcal{R}$ and $\Lambda-\mathcal{R}$, and their CFI values are generally nondegenerate. Figure~\ref{fig4:supp} tracks the branches associated with the largest positive and negative fractional variance changes for nonzero detuning, and their abrupt switching is explained in the next section.

\subsubsection{Supermode Representation}
Since the two photocurrent contributions are identical, the full sideband-pair CFI can be optimized using only the $+$ supermode. Defining $\mathbf{u}=\left(u_1, u_2 \right)^T=\left(\cos\ph, \sin\frac{\phi}{2} \right)^T$, the CFI is
\fla{\mathcal{F}_{\rm pair}(\omega,\phi)=\frac{1}{2}\left(\frac{\mathbf{u}^T \Sp'\mathbf{u}}{\mathbf{u}^T \Sp\mathbf{u}}\right)^2+\frac{1}{2}\left(\frac{\mathbf{u}^T\mathbf{J}^T \Sm'\mathbf{J}\mathbf{u}}{\mathbf{u}^T \mathbf{J}^T\Sm \mathbf{J}\mathbf{u}}\right)^2=\left(\frac{\mathbf{u}^T \Sp'\mathbf{u}}{\mathbf{u}^T \Sp\mathbf{u}}\right)^2.}
Since the quotient is invariant under rescaling $\mathbf u$, constrained optimization gives the generalized eigenvalue equation $\boldsymbol{\Sigma}_{\pm}'\mathbf{u}=\alpha_{1,2}\boldsymbol{\Sigma}_{\pm}\mathbf{u}$, with $\mathcal{F}_{\rm pair}^{\rm opt}(\omega)=\max(\alpha_1^2,\alpha_2^2)$ and optimal quadrature angle $\phi_{\rm opt}=2\operatorname{atan2}(u_2,u_1)$. For $\gamma=0$, $\det\Sp=1$, so $\alpha_1+\alpha_2=\operatorname{Tr}(\Sp^{-1}\Sp')=\partial_{\delta\omega}\ln\det\Sp=0$, and hence $\alpha_1=-\alpha_2\equiv\alpha$. Using Eq.~\eqref{eq::sm-qfi} gives
\fla{\Ipair=2\mathcal{I}_+=\frac{1}{2}\operatorname{Tr}[\Sp^{-1}\Sp'\Sp^{-1}\Sp']=\alpha^2,}
Thus, in the fully monitored OPO, measuring the optimal frequency-dependent
quadrature saturates the QFI of every sideband pair. As in the main text,
this frequency dependence can be implemented by spectrally phase shifting
the OPO output before detection with a fixed-phase monochromatic LO. In the presence of intrinsic loss, the output state is mixed, so instead we have 
\fla{2\mathcal{I}_+=\frac{\operatorname{Tr}[\Sp^{-1}\Sp'\Sp^{-1}\Sp']}{1+\mu^2}+4\frac{(\mu')^2}{1-\mu^4}=\frac{\alpha_1^2+\alpha_2^2}{1+\mu^2}+\frac{\mu^2(\alpha_1+\alpha_2)^2}{1-\mu^4},}
where we used $\partial_{\delta\omega}\mu=-\frac{\mu}{2}\partial_{\delta\omega}\ln{\det{\Sp}}$.
The difference between the QFI and CFI is then
$2\mathcal{I}_+-\mathcal{F}_{\rm pair}^{\rm opt}(\omega)=\frac{(\alpha_2+\mu^2\alpha_1)^2}{1-\mu^4}$,
assuming $|\alpha_1|>|\alpha_2|$ ordering, which is generally nonzero. The same reasoning applies at $\omega=0$, with both QFI and CFI equal to half the corresponding sideband-pair values.

\begin{figure}
    \centering
    \includegraphics[width=0.6\linewidth]{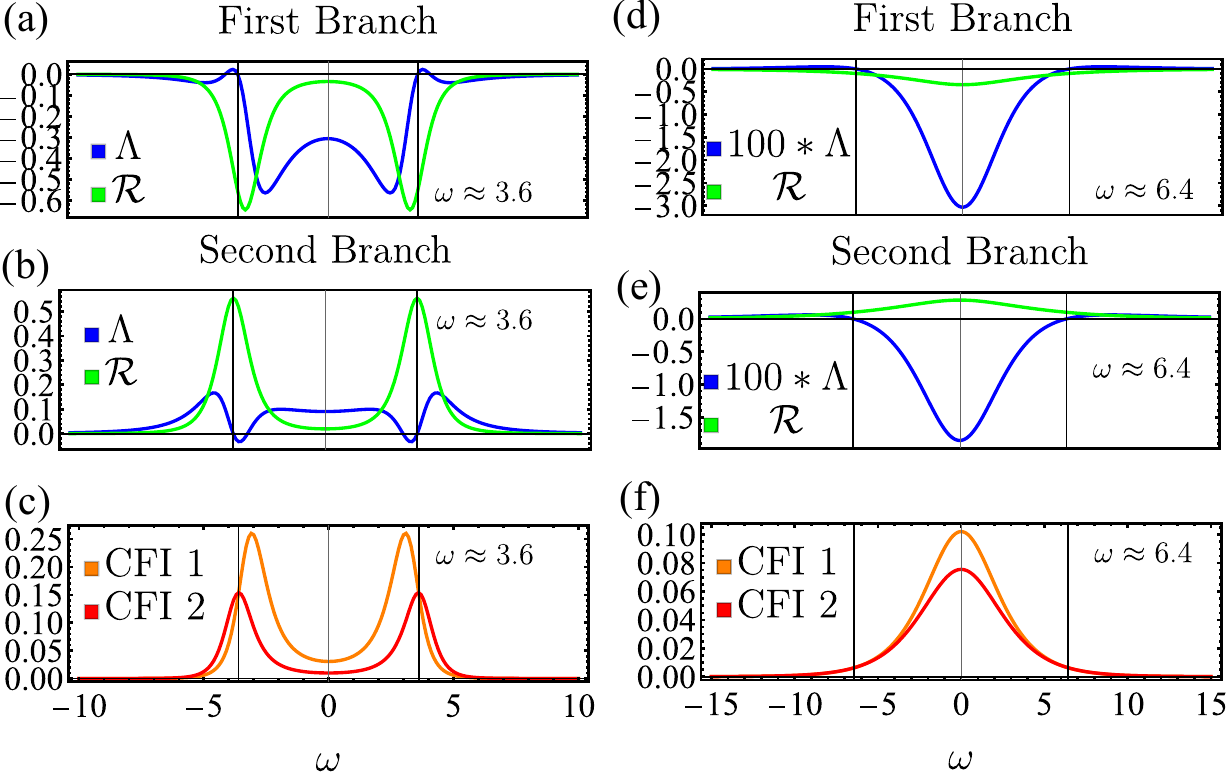}
    \vspace{-0.3cm}
    \caption{\textbf{Eigenvalue and rotation contributions to the fractional change in variance.} (a)--(c) Parameters used for the double-peaked QFI spectrum, $\delta\omega=4,\; \xi=2,\;\gamma=\kappa=1$. (d)--(f) A single-peaked QFI example with $\delta\omega=1,\; \xi=4,\;\gamma=5,\; \kappa=10$. First branch refers to optimal quadrature angle for central frequencies, while second branch refers to optimal quadrature angle solution for outer frequencies. Evaluating the CFI at the first and second branches gives CFI 1 and CFI 2. In both cases, when $\Lambda$ for the initially dominant solution branch (first branch) crosses zero (vertical black line), the globally optimal quadrature angle switches to the second branch, reflected in the crossing of the corresponding CFI maxima.}
    \label{fig5:supp}
\end{figure}

For the OPO, the two generalized eigenvalues have opposite signs and correspond to the largest positive and negative fractional changes in variance, and both eigendirections are local CFI maxima. Writing $\mathcal{F}_{\rm pair}(\omega,\phi)=R^2$, with $R=\mathbf v^T\mathbf K\mathbf v/(\mathbf v^T\mathbf v)$, $\mathbf K=\Sp^{-1/2}\Sp'\Sp^{-1/2}$, and $\mathbf v=\Sp^{1/2}\mathbf u$, the Rayleigh quotient $R$ is bounded by the eigenvalues $\alpha_{1,2}$ of the real-symmetric matrix $\mathbf{K}$. Since $\alpha_{1,2}$ have opposite signs, moving away from either eigendirection moves $R$ toward the interior of $[\alpha_1,\alpha_2]$, reducing $|R|$ and hence the CFI. In Fig.~\ref{fig5:supp}, we track both branches, whose dominance switches at $\Omega_*=\frac12\sqrt{\Gamma^2+4(\delta\omega^2-\xi^2)}$. At $\Omega_*$, $\lambda_s'=\lambda_a'=0$, so $\Lambda=0$. Since the mirror-related branches satisfy $\mathcal{F}_{\rm pair}(\omega,\phi)(\chi)-\mathcal{F}_{\rm pair}(\omega,\phi)(-\chi)=4\Lambda\mathcal{R}/V^2$ with $\chi=\phi+\theta$, they are degenerate at this frequency. Moreover, $\left.\partial_\omega\Lambda\right|_{\Omega_*}\ne0$ for finite $\gamma,\kappa$, so $\Lambda$ changes sign across $\Omega_*$ when $\delta\omega\ne 0$. With $\mathcal{R}\ne0$ at the maxima, their relative heights reverse, producing the branch switching shown in Fig.~\ref{fig5:supp}.

\subsubsection{Classical Fisher Information Scaling with Photon Number}
We consider how the classical Fisher information scales as threshold is approached. Optimized homodyne detection at each $\omega$ extracts $\mathcal{F}_{\rm pair}^{\rm opt}(\omega)=\max(\alpha_1^2,\alpha_2^2)$, where $\alpha_{1,2}$ are the eigenvalues of $\mathbf{K}=\Sp^{-1/2}\Sp'\Sp^{-1/2}$. It is easier to do the analysis in the principal axis basis with $\tilde{\mathbf{K}}=\mathbf{M} \mathbf{K}\mathbf{M}^T=\begin{pmatrix}
    A & C \\ C & B
\end{pmatrix}$, where $A=\lambda_a'/\lambda_a$, $B=\lambda_s'/\lambda_s$, and $C=\frac{\theta'}{2}\frac{\lambda_s-\lambda_a}{\sqrt{\lambda_a \lambda_s}}$. We first take $\bar n\to\infty$ at fixed positive $\kappa,\gamma,\delta\omega$ and fixed $z$, and subsequently consider additional parameter limits subject to the validity conditions below. After some algebraic manipulations, the leading orders of each coefficient are found to be $A\sim -\frac{8\delta\omega \xi_c^2}{\xi_c^4+4\Gamma^2 z^2}\nbar$, $B\sim\frac{\delta\omega \kappa}{2\Gamma^2 \gamma \nbar}$, $C\sim-\frac{4\nbar \xi_c \sqrt{\kappa/\gamma}}{\sqrt{\xi_c^4+4 \Gamma^2 z^2}}$, and $C^2\sim\frac{16\kappa \xi_c^2}{\gamma(\xi_c^4+4 \Gamma^2 z^2)}\nbar^2$. The eigenvalues are evaluated as $\alpha_{\pm}\sim \frac{A\pm \sqrt{A^2+4C^2}}{2}$, where the term proportional to $B$ has been dropped. For $\delta\omega>0$, $A<0$, so that $|\alpha_-|>|\alpha_+|$ in the critical region, and the asymptotic information extracted by homodyne is
\fla{\mathcal{F}_{\rm pair}^{\rm opt}\left(\frac{z}{\nbar}\right)\sim\alpha_-^2\sim \frac{16 \nbar^2 \xi_c^4}{(\xi_c^4+4 \Gamma^2 z^2)^2}\left[\delta\omega+\sqrt{\delta\omega^2+\frac{\kappa(\xi_c^4+4\Gamma^2 z^2)}{\gamma \xi_c^2}}\right]^2.}
Also note that for fixed $\kappa,\gamma>0$ as $\nbar\to \infty$, the purity in the critical region goes to zero, so that $\mathcal{I}_{\rm pair} \to \alpha_{+}^2+\alpha_{-}^2\sim A^2+2C^2$, which reproduces Eq.~\eqref{eq::Ipairzn} after substituting the asymptotic expressions for $A,C$. 
Defining $\dot{\mathcal{F}}^{\rm opt}=\frac{1}{2\pi} \int_0^\infty d\omega \mathcal{F}_{\rm pair}^{\rm opt}(\omega)$, analogously to $\dot{\mathcal{I}}_{\rm out}$, direct integration gives 
\fla{\eta=\frac{\dot{\mathcal{F}}^{\rm opt}}{\dot{\mathcal{I}}_{\rm out}}=\frac{1}{2}+\frac{1}{\pi}\arctan \left(\frac{\delta\omega/\xi_c}{\sqrt{\kappa/\gamma}}\right)+\frac{\delta\omega/\xi_c \sqrt{\kappa/\gamma}}{\pi (\delta\omega^2/\xi_c^2+\kappa/\gamma)}.}
The extraction efficiency is an increasing function of $\frac{\delta\omega/\xi_c}{\sqrt{\kappa/\gamma}}$, reaching unity efficiency when this quantity goes to infinity. In the overcoupled ($\kappa/\gamma \to \infty$) near-threshold bound saturating regime, $\eta$ approaches $\frac{1}{2}$, regardless of how $0\le \frac{\delta\omega}{\xi_c}\le 1$ varies. This can also be understood by considering the ratio $\frac{A^2}{C^2}=\frac{4\gamma}{\kappa}\frac{\delta\omega^2\xi_c^2}{\xi_c^4+4\Gamma^2 z^2}\le\frac{4\gamma}{\kappa}$, which tends to zero as $\frac{\kappa}{\gamma} \to \infty$, so that $C^2\gg A^2$, allowing to approximate $\alpha_{\pm}\sim \pm |C|$. Therefore, the two generalized eigendirections contribute equally to the leading QFI, while homodyne detection can only select one of them. Conversely, in the far-detuned limit, $\eta\to \frac{1}{2}+\frac{1}{\pi} \arctan \sqrt{\frac{\gamma}{\kappa}}+\frac{\sqrt{\kappa/\gamma}}{\pi (1+\kappa/\gamma)}$, so that $\eta\to 1$ can be approached if $\frac{\kappa}{\gamma} \to 0$. Note that $\frac{\delta\omega}{\Gamma}\to \infty$ needs to be maintained simultaneously with $\frac{\kappa}{\gamma} \to 0$ to ensure both the QFI bound saturation and near-unity $\eta$. Starting in the far-detuned regime, one may decrease $\kappa$ at fixed $\gamma$, so that the QFI bound coefficient itself does not change. Indeed, if $\kappa \to 0$ at fixed $\gamma$ and $\delta\omega\ne 0$, $C^2/A^2\to 0$, so that nearly all QFI is concentrated in the eigendirection corresponding to $\alpha_-\sim A$, while $\alpha_+\sim 0$, so that homodyne can just select the most informative one.
Note that the assumptions underlying the asymptotic QFI treatment also need to be respected as $\kappa,\delta\omega$ are varied, i.e., $Q(\omega)\ll 2 \gamma \kappa \xi_c^2$ in the critical bandwidth. Setting $\omega=\Delta\omega_{\rm QFI} x$ where $x=\mathcal{O}(1)$ within the critical bandwidth then imposes $\frac{\xi_c^2}{\gamma\kappa\nbar^2}\ll 1$.

We now consider the behaviour of the optimal quadrature in the critical bandwidth for $\delta\omega>0$. The general form of the eigendirection of $\tilde{\mathbf{K}}$ corresponding to $\alpha_-$ can be written as $\tilde{v}=(\cos \beta,\sin \beta)^T$. Finding an inner product with its orthogonal vector $(-\sin \beta,\cos \beta)$ imposes $\tan 2\beta=\frac{\sqrt{\kappa/\gamma}}{\delta\omega/\xi_c}\sqrt{1+x^2}>0$, where $x=\frac{2\Gamma z}{\xi_c^2}$, and since $\tan\beta=C/\alpha_->0$, we can fix $0<\beta<\frac{\pi}{4}$. Transforming to the corresponding quadrature angle $\tilde{\phi}$
relative to the principal-axis basis gives $\tan \tilde{\phi}=\sqrt{\frac{\lambda_a}{\lambda_s}}\tan \beta$, so that the optimal eigendirection is $(\cos \tilde{\phi},\sin{\tilde{\phi}})^T$. Solving for $\tan\beta$:
\fla{\tan \tilde{\phi}(x)=\frac{4\Gamma \nbar}{\xi_c}\frac{\kappa/\gamma}{(\delta\omega/\xi_c)\left(1+\sqrt{1+\frac{\kappa/\gamma}{\delta\omega^2/\xi_c^2}(1+x^2)}\right)}.}
In the limit $\frac{\sqrt{\kappa/\gamma}}{\delta\omega/\xi_c}\to 0$ for fixed $x=\mathcal{O}(1)$ within the critical bandwidth, the optimal quadrature directions have the same leading asymptotic slope in the principal-axis basis, since $\frac{\tan \tilde{\phi}(x)}{\tan \tilde{\phi}(0)}\sim 1-\frac{\kappa/\gamma}{4 \delta\omega^2/\xi_c^2}x^2\to 1$. Along the path considered in the main text, the principal-axis directions themselves also have the same leading-order asymptotic orientation across the critical region $\frac{\tan \theta(x)}{\tan\theta(0)}\sim 1-\frac{\xi_c^4}{2\Gamma^4 \nbar^2}x^2\to 1$. Consequently, the optimal quadrature directions in the laboratory phase-space basis agree increasingly well under $\frac{\sqrt{\kappa/\gamma}}{\delta\omega/\xi_c}\to 0$ and $\frac{\xi_c^4}{2\Gamma^4 \nbar^2}=\frac{(\delta\omega^2/\Gamma^2+1/4)^2}{2 \nbar^2}\to 0$, equivalent to $\delta\omega/\Gamma$ growing slower than $\sqrt{\nbar}$, which is satisfied along far-detuned critical paths at fixed finite $\Gamma$.

\bibliography{references_abbreviated}